\documentclass[reqno,12pt]{amsart}

\usepackage{geometry,hyperref}
\usepackage[none]{hyphenat}
\usepackage{amssymb}
\usepackage{bm,enumerate}
\usepackage{graphicx,xcolor,xspace,booktabs,caption,makecell}
\usepackage{subfig,float}
\usepackage{enumitem}
\usepackage{etoolbox}

\makeatletter
\patchcmd{\@settitle}{\uppercasenonmath\@title}{\large}{}{}
\makeatother

\newcommand{\bx}{\bm{x}}
\newcommand{\bv}{\bm{v}}
\newcommand{\bu}{\bm{u}}
\newcommand{\bU}{\bm{U}}
\newcommand{\bF}{\bm{F}}
\newcommand{\bI}{\mathbf{I}}
\newcommand{\bW}{\bm{W}}
\newcommand{\bG}{\bm{G}}

\newcommand{\bw}{\bm{w}}

\newcommand{\bR}{\bm{R}}

\newcommand{\bmm}{\bm{m}}
\newcommand{\bphi}{\bm{\phi}}
\newcommand{\bM}{\bm{M}}
\newcommand{\veps}{\varepsilon}

\newcommand{\tF}{\tilde{\bm{F}}}
\newcommand{\tG}{\tilde{\bm{G}}}
\newcommand{\Gal}{\text{Gal}}

\newcommand{\dt}{\partial_t}
\newcommand{\dx}{\nabla_{\bm{x}}}
\newcommand{\R}{\mathbb{R}}
\newcommand{\E}{\mathbb{E}}
\newcommand{\cD}{\mathcal{D}}

\newcommand\taska{\textit{Task Wave}\xspace}
\newcommand\taskb{\textit{Task Mix}\xspace}
\newcommand\taskc{\textit{Task MixInTransition}\xspace}
\newcommand\secenc{\textit{Finding Generalized Moments}\xspace}
\newcommand\secclosure{\textit{Learning Moment Closure}\xspace}
\newcommand\secgalilean{\textit{Symmetries and Galilean Invariant Moments}\xspace}
\newcommand\secdirectconv{\textit{An Alternative Direct Machine Learning Strategy}\xspace}

\newcommand*\diff{\mathop{}\!\mathrm{d}}

\newcommand{\transpose}{^{\operatorname{T}}}

\calclayout
\begin{document}

\title[Uniformly Accurate Hydrodynamic Models]{Uniformly Accurate Machine Learning-Based Hydrodynamic Models for Kinetic Equations}

\author[Jiequn Han \and Chao Ma \and Zheng Ma \and Weinan E]
{Jiequn Han$^{1,*}$ \and Chao Ma$^2$ \and Zheng Ma$^3$ \and Weinan E$^{1,2,4}$}
\thanks{$^*$To whom correspondence should be addressed. E-mail: jiequnh@princeton.edu}
\makebox[0pt]{\phantom{\footnotemark}}\footnotetext{Department of Mathematics, Princeton University}
\makebox[0pt]{\phantom{\footnotemark}}\footnotetext{Program in Applied and Computational Mathematics, Princeton University}
\makebox[0pt]{\phantom{\footnotemark}}\footnotetext{Department of Mathematics, Purdue University}
\makebox[0pt]{\phantom{\footnotemark}}\footnotetext{Beijing Institute of Big Data Research}

\begin{abstract}
A new framework is introduced for constructing interpretable and truly reliable reduced models for multiscale problems in situations without scale separation. Hydrodynamic approximation to the kinetic equation is used as an example to illustrate the main steps and issues involved. To this end, a set of generalized moments are constructed first to optimally represent the underlying velocity distribution. The well-known closure problem is then solved with the aim of best capturing the associated dynamics of the kinetic equation. The issue of physical constraints such as Galilean invariance is addressed and an active learning procedure is introduced to help ensure that the dataset used is representative enough. The reduced system takes the form of a conventional moment system and works regardless of the numerical discretization used. Numerical results are presented for the BGK (Bhatnagar--Gross--Krook) model and binary collision of Maxwell molecules. We demonstrate that the reduced model achieves a uniform accuracy in a wide range of Knudsen numbers spanning from the hydrodynamic limit to free molecular flow.
\end{abstract}
\maketitle

In scientific modeling, we often encounter the following dilemma: 
We are interested in modeling some macro-scale phenomenon but we only have a reliable model at some micro scale and the
 micro-scale model is too detailed and too expensive for practical use.
The idea of multiscale modeling is to develop models or algorithms 
 that can efficiently capture the macro-scale behavior of the system, but use only the 
micro-scale model  (see \cite{weinan2007heterogeneous,weinan2011principles} for a review).
This program has been quite successful for situations  in which there is a clear gap between the 
active scales in  the macro- and micro-scale processes. However, 
for problems that do not exhibit scale separation, 
we still lack effective
tools that can be used to discover the relevant variables and find accurate approximations for the dynamics of these relevant variables.

Machine learning, the second topic that we are concerned with here, is equally broad and important.
We are particularly interested in  developing physical models using machine learning.
There are significant differences between this and  traditional machine learning tasks such as the ones in computer vision and data analytics.  One is that in the current setting, instead of being given the data, the data is generated using the
micro-scale model. The good news is that in principle we can generate an unlimited amount of data. The
bad news is that often times the process of generating data is expensive. Therefore it is an important issue to find a
 data set that is as small as possible and yet representative enough. 
The second aspect  is that we have to be concerned with physical constraints such as 
symmetry, invariance, and the laws of nature. We have to make a choice between enforcing these constraints explicitly,
or enforcing them approximately as a byproduct of accurately modeling the underlying physical process.
The third is the interpretability issue. Machine learning models are often black-box in nature.
But as a physical model that we can rely on, just as the Schr\"odinger equation in quantum mechanics or Navier--Stokes equation
in fluid mechanics, it cannot just be  a black-box completely.
Some of these issues have been studied in \cite{han2017deep,zhang2018deep,zhang2019active} in the context of modeling 
inter-atomic force fields. 

The work presented in \cite{han2017deep,zhang2018deep,zhang2019active} is limited to the context of learning some functions.
This paper examines the aforementioned issues systematically in the context of developing new
dynamical models for multiscale problems.
As a concrete example, we will study the problem where the micro-scale model is the kinetic equation, and the macro-scale model is the hydrodynamic model.  Besides being a very representative problem in multiscale modeling,
this example  is also of great practical interest in its own right.
Kinetic equations %
have found applications in modeling rarefied gases, plasmas, micro-flow, semiconductor devices, radiative transfer, complex fluids, and so on. 
From a computational point of view,  kinetic equations are costly to solve due to the dimensionality of
the problem and the multi-dimensional integral in the collision term. 
As a result, there has been a long history of interest  in reducing kinetic models to hydrodynamic models,
going back at least to the work of Grad \cite{grad1949kinetic}. Our work is in the same spirit.

\begin{sloppypar}
A crucial dimensionless number that influences the behavior of the solutions of the kinetic equations is the Knudsen number, defined as the ratio between the mean free path of a particle and the typical macroscopic length scale of interest.
When the Knudsen number is small, the velocity distribution stays close to  local Maxwellians.
This is the so-called hydrodynamic regime. In this regime, it is possible to derive hydrodynamic models for some selected macroscopic fluid variables (typically the mass, momentum, and energy densities) such as the Euler equations or the Navier--Stokes--Fourier equations \cite{bardos1993fluid,levermore1996moment}. 
These hydrodynamic models are not only much less costly to solve, but also much easier to use to explain experimental observations. They are often in the form of conservation laws with some conserved quantities and fluxes.
However, when the Knudsen number is no longer small, the hydrodynamic models break down. Direct simulation Monte Carlo (DSMC) method ~\cite{bird1994molecular,alexander1997direct,pareschi2001introduction} 
becomes a more preferred choice.
DSMC works well in the collision-less limit when the Knudsen number is large, but the associated computational cost becomes prohibitive  
when approaching the hydrodynamic regime due to the high collision rate.
The significant difference between these two types of approaches creates  a problem when modeling  transitional flows
in which the Knudsen number varies drastically. 
\end{sloppypar}

Significant effort has been devoted to developing generalized hydrodynamic models as a replacement of the kinetic equation in different regimes, with limited success.
These generalized models can be put into two categories.  
The first  are direct extensions of the Navier--Stokes--Fourier equations. 
In these models, no new variables are introduced, but new derivative terms are added.
One example is the Burnett equation \cite{burnett1935distribution}.
The second are the moment equations.  In this case additional moments are introduced and their dynamic
equations are derived  through the process of ``moment closure'' using the
kinetic equation.  The most well-known example is Grad's 13-moment equations \cite{grad1949kinetic}. 
Much effort has gone into making sure that such moment equations are mathematically well-posed and respect the most important
physical constraints, see  \cite{levermore1996moment,cai2014globally,cai2015framework}.
In both approaches, one has to introduce some uncontrolled  truncation in order to obtain a 
closed system of partial differential equations (PDE).

To develop machine learning-based models, one can also proceed along these two separate directions.
One can stay with the original set of hydrodynamic variables (mass, momentum and energy densities), but introduce  corrections to the fluxes as functionals of the local (in space) instantaneous hydrodynamic variables.
This approach is relatively straightforward to implement through convolutional neural networks, and it does give reasonably satisfactory results.
However, the accuracy of such a model is limited beforehand by the choice of the hydrodynamic variables. Moreover, 
the resulting model  depends heavily on the specific spatial discretization used to find the data and train the model,
and it is difficult to obtain a continuous PDE-like model using this approach.
In other words, the result is more like a specific numerical algorithm rather than a new reliable physical model.
In addition, some of the issues that we mentioned above, which  are important for more general multiscale modeling problems, 
do not arise. This means that as an illustrative example for a general class of problems, it has a limited value.
For these reasons, the majority of our efforts is devoted to the second category of methods:  the moment equations.
Specifically, we revisit the conventional Grad-type moments and explore the possibility of learning new generalized moments, not necessarily polynomials, to represent the velocity distribution function.
We also explore the possibility of learning  accurately the  dynamical model, instead of resorting to ad hoc closure assumptions.
In addition, we develop an active learning procedure in order to ensure that the data set used to generate these models are representative enough.

Below we use the Boltzmann equation to explain the technical details of our approach.
The algorithms are implemented for the BGK (Bhatnagar--Gross--Krook) model \cite{bhatnagar1954model} and  the Boltzmann
equation modeling binary collision of Maxwell molecules.
We work with a wide range of Knudsen numbers ranging from $10^{-3}$ to 10 and different initial conditions. 
It is easy to see that the methodology developed here is applicable to a much wider range of models,
some of which are under current investigation.

There has been quite some activities recently on applying machine learning, especially deep learning to study dynamical systems, including representing the physical quantities described by PDEs with physics-informed neural networks \cite{raissi2018hidden,raissi2019physics}, uncovering
the underlying hidden PDE models with convolutional kernels~\cite{long2017pde,long2018pde} or sparse identification of nonlinear dynamical systems  method~\cite{rudy2019data,champion2019discovery}, predicting reduced dynamics with memory effects using long short-term memory recurrent neural networks~\cite{ma2018model,vlachas2018data}, 
calibrating the Reynolds stress in the Reynolds averaged Navier--Stokes equation~\cite{ling2016reynolds,wang2017physics}, and so on. Machine learning techniques have also been leveraged to find  proper coordinates for dynamical systems in the spirit of dimensionality reduction~\cite{li2017extended,takeishi2017learning,lusch2018deep,champion2019data}. Despite the widespread interests in machine learning of dynamical systems, the examples presented in the literature are mainly ODE models or PDE models in which the ground truth is known. This paper stands in contrast to previous works with the aim of learning, from the original micro-scale model, new reduced PDEs at the macro scale accurately.

\section{Preliminaries}
Let $D$ be the spatial dimension and $f=f(\bx, \bv, t)\colon \R^D \times \R^D \times \R \rightarrow \R$ be the one-particle phase space probability density.
Consider the following initial value problem of the Boltzmann equation in the absence of external forces
\begin{equation}
\label{eq:Boltzmann}
  \dt f + \bv\cdot\dx f = \frac{1}{\veps} Q(f), \quad f(\bx,\bv, 0) = f_0(\bx, \bv),
\end{equation}
where $\varepsilon$ is the dimensionless Knudsen number and $Q$ is the collision operator.
Two examples of the collision operator are considered in this paper.
The first example is the BGK model
\begin{equation}
\label{eq:Q_bgk}
    Q(f)(\bv) = f_M(\bv) - f(\bv).
\end{equation}
Here $f_M$ is the local Maxwellian distribution function, sometimes also called the local equilibrium,
\begin{equation}
\label{eq:Maxwell}
  f_M(\bv) = \frac{\rho}{(2\pi T)^{\frac{D}{2}}}\exp\left(-\frac{|\bv - \bu|^2}{2T}\right),
\end{equation}
where $\rho$, $\bu$ and $T$ are the density, bulk velocity and temperature fields. They are related to the moments of $f$ through
\begin{equation}
\label{eq:def_rhomE}
  \rho = \int_{\R^D} f\diff\bv, \quad \bu = \frac{1}{\rho}\int_{\R^D} f\bv\diff\bv, \quad T = \frac{1}{D\rho}\int_{\R^D} f |\bv - \bu|^2\diff\bv.
\end{equation}
In above the dependence on the location and time has been dropped for clarity of the notation.
The second example considered is the model for binary collision of Maxwell molecules in 2-D
\begin{equation}
\label{eq:Q_maxwell}
    Q(f)(\bv) = \int_{\mathbb{R}^2}\int_{\mathbb{S}^1}\frac{1}{2\pi}(f(\bv')f(\bv_*')-f(\bv)f(\bv_*))\diff \sigma \diff \bv_*.
\end{equation}
Here $(\bv, \bv_*)$ and $(\bv', \bv_*')$ are the pairs of pre-collision and post-collision velocities, related by
\begin{equation*}
    \left\{
  \begin{aligned}
     & \bv' = \frac{\bv+ \bv_*}2 +  \frac{|\bv - \bv_*|}{2}\sigma, \\
     & \bv_*' = \frac{\bv+ \bv_*}2 -  \frac{|\bv - \bv_*|}{2}\sigma,
  \end{aligned}
  \right.
\end{equation*}
with $\sigma\in\mathbb{S}^1$.
In general $Q$ satisfies the conditions
\begin{equation}
\label{collision_consv}
  \int_{\R^D} Q(f)(\bv)\bphi(\bv)\diff\bv = \bm{0}, \quad \bphi(\bv) = (1, \bv, |\bv|^2/2)\transpose.
\end{equation}
These ensure that mass, momentum and energy are conserved during the evolution.

It can be shown that when $\varepsilon \ll 1$, $f$ stays close to the local Maxwellian distributions \cite{cercignani1975theory,chapman1990mathematical,nishida1978fluid,caflisch1980fluid,bouchut2000kinetic}.
Taking the moments $\langle\cdot, \bphi\rangle := \int \cdot\bphi(\bv)\diff\bv$ on both sides of the Boltzmann equation \eqref{eq:Boltzmann} with $\bphi$ in \eqref{collision_consv} and replacing $f$ by $f_M$,
one obtains the compressible Euler equations:
\begin{equation}
\label{euler_full}
  \left\{
  \begin{aligned}
     & \dt \rho + \dx\cdot (\rho \bu) = 0,                     \\
     & \dt (\rho\bu) + \dx\cdot (\rho \bu\otimes\bu + p\bI) = 0, \\
     & \dt E + \dx\cdot ((E + p)\bu) = 0,
  \end{aligned}
  \right.
\end{equation}
where $p = \rho T$ is the pressure and $E = \frac{1}{2}\rho |\bu|^2 + \frac{D}{2}\rho T$ is the total energy.
Let
\begin{equation}
\label{eq:euler_F}
  \bU = \begin{pmatrix}
  \rho \\
  \rho\bu \\
  E
  \end{pmatrix},
  \quad \bF_{\text{Euler}}(\bU) = \begin{pmatrix}
  \rho\bu\transpose \\
  \rho \bu\otimes\bu + p\bI\\
  (E + p)\bu\transpose
  \end{pmatrix},
\end{equation}
we can rewrite the Euler equations in a succinct conservation form
\begin{equation}\label{euler}
  \dt \bU + \dx\cdot \bF_{\text{Euler}}(\bU) = 0.
\end{equation}
For larger values of $\veps$ one would like to use the moment method to obtain similar systems of PDEs that serve as an approximation to the Boltzmann equation. 
To this end, 
one starts with  a linear space $\mathbb{M}$ of functions of $\bv$ (usually chosen to be polynomials) that include the components of $\bphi$ in~\eqref{collision_consv}.
Denote by  $\bmm$, a vector whose components form a basis of $\mathbb{M}$. Then the moments $\bM$, as functions of $\bx$ and $t$,
 are defined as $\int_{\R^D}f\bmm(\bv)\diff \bv$.  From the Boltzmann equation, one obtains
\begin{equation}
    \label{eq:integral_Boltz}
    \partial_t\bM + \dx \cdot \int_{\R^D}f\bmm(\bv)\bv\transpose \diff \bv = \int_{\R^D}\frac1\veps Q(f)\bmm(\bv)\diff \bv.
\end{equation}
 The challenge is  to approximate the two integral terms in~\eqref{eq:integral_Boltz} as functions of $\bM$ in order to obtain
 a closed system.
 This is the well-known ``moment closure problem'' and it is at this stage that various uncontrolled approximations  are introduced.
  In any case, once this is done one obtains a closed system of the form
\begin{equation}
\label{moment_dynamics}
  \left\{
  \begin{aligned}
     & \dt \bU + \dx\cdot \bF(\bU, \bW) = 0, \\
     & \dt \bW + \dx\cdot \bG(\bU, \bW) = \frac{1}{\veps}\bR(\bU, \bW).
  \end{aligned}
  \right.
\end{equation}
Here the moments $\bM$ are decomposed as $\bM=(\bU, \bW)\transpose$, where $\bU \in \R^{D+2}$ denotes the conserved quantities including mass, momentum and energy densities, and $\bW \in \R^{M}$ are the extra new moments. The basis functions $\bmm(\bv)$ are decomposed accordingly as $\bmm(\bv)=(\bphi(\bv), \bw(\bv))\transpose$.
The initial condition of~\eqref{moment_dynamics} is determined by the one of the Boltzmann equation~\eqref{eq:Boltzmann}: 
$$\bU(\bx, 0)=\int_{\R^D}f_0(\bx, \bv)\bphi(\bv)\diff\bv, \quad \bW(\bx, 0)=\int_{\R^D}f_0(\bx, \bv)\bw(\bv)\diff\bv.$$
Similarly, the term $1/\veps$ in~\eqref{moment_dynamics} is inherited directly from~\eqref{eq:Boltzmann}.

\section{Machine Learning-Based Moment System}
\begin{figure}[ht]
    \centering
    \includegraphics[width=0.95\textwidth]{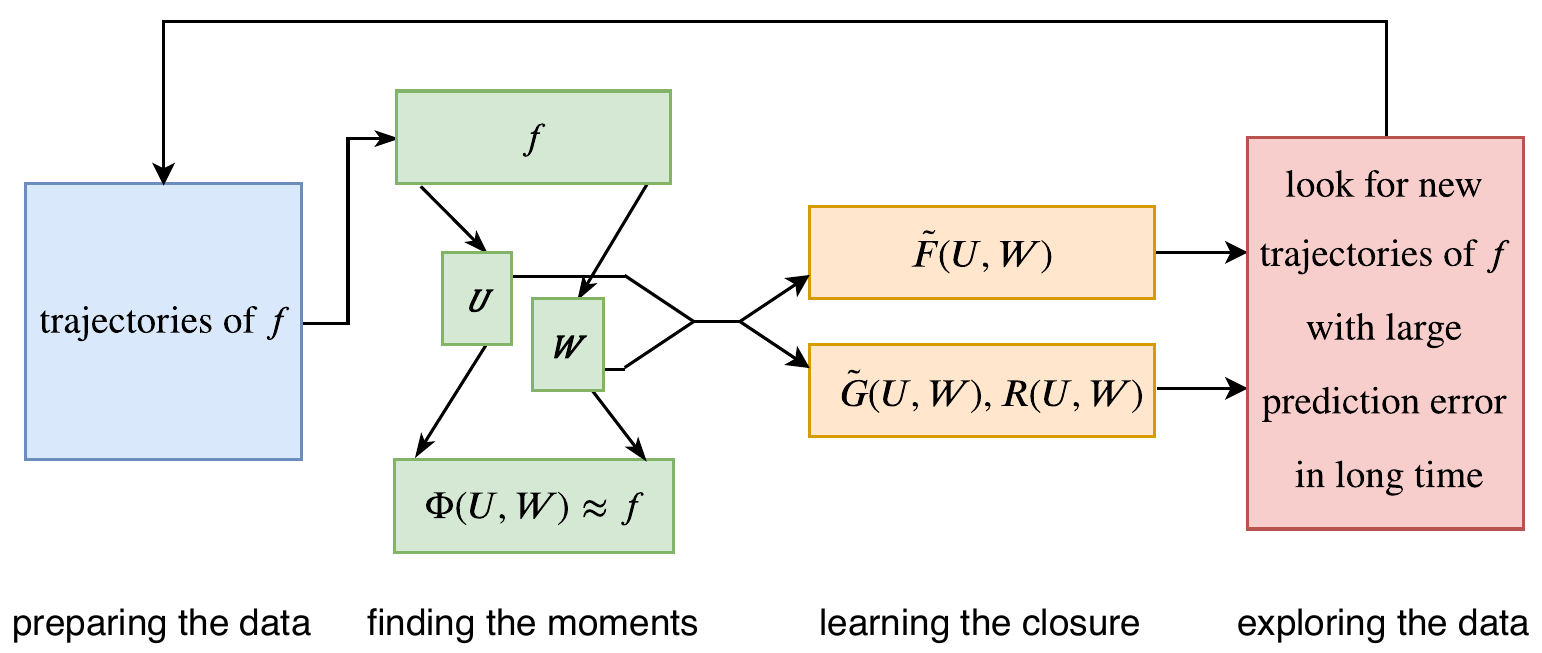}
    \caption{Schematic diagram of the machine learning-based moment method}
    \label{fig:schematic}
\end{figure}

\begin{sloppypar}
We are interested in approximating a family of kinetic problems, in which the Knudsen number may span from the hydrodynamic regime ($\veps\ll 1$) to the free molecular regime ($\veps \sim 10 $), and the initial conditions are sampled from a wide distribution of profiles. 
Fig.~\ref{fig:schematic} presents a schematic diagram of the framework for the machine learning-based moment method.
Below we first describe  the method for finding generalized moments and addressing the moment closure problem. Then we discuss how to build the data set $\cD$ incrementally to achieve efficient data exploration.
\end{sloppypar}

\subsection{Finding Generalized Moments}
\label{sec:learn_mom}
For this discussion, we can neglect the dependence of $f$ on $(\bx, t)$ and view $f$ as a function of the velocity $\bv$ only. We consider two different ways of finding additional moments $\bW$. One is using the conventional Grad-type moments and the other  is based on the autoencoder.
The conventional moment system, such as Grad's moment system, starts with the Hermite expansion of $f$
\begin{equation}
\label{eq:f_expansion}
    f(\bv) = \sum_{\bm{\alpha}\in\mathbb{N}^D}f_{\bm{\alpha}}\mathcal{H}_{T,\bm{\alpha}}\left(\frac{\bv-\bu}{\sqrt{T}}\right),
\end{equation}
where $\bm{\alpha}=(\alpha_1,\dots,\alpha_D)$ is a $D$-dimensional multi-index. Here 
the basis functions are defined as
$$
  \mathcal{H}_{T, \bm{\alpha}}(\bv) = \frac{1}{(2\pi)^\frac{D}{2}}\prod_{j=1}^D T^{-\frac{\alpha_j+1}{2}} He_{\alpha_j}(v_j)\exp(-v_j^2/2),
$$
and $He_k(\cdot)$ is the $k$-th order Hermite polynomial. 
Using the orthogonality of Hermite polynomials, one can easily express $f_{\bm \alpha}$ as certain moment of $f$.
If $f$ is at the local equilibrium, the only non-zero coefficient is $f_{\bm{0}}=\rho$. In general, one can truncate the Hermite expansion at an order $L$ and select those coefficients $f_{\bm \alpha}$ with $0<|\bm\alpha|\leq L$ to construct the additional moments $\bW_{\text{Herm}}$. 

The aforementioned Grad-type moments are based on the rationale that the truncated Hermite series is an effective approximation to $f$ near the local equilibrium. In a broader sense, we can use an autoencoder~\cite{hinton2006reducing,goodfellow2016deep} to seek the generalized moments $\bW_{\text{Enc}}$ without relying on such domain knowledge.
More concretely, we wish to find an encoder $\Psi$ that maps $f$ to generalized moments $\bW_{\text{Enc}} \in \R^M$ and a decoder $\Phi$ that recovers the original $f$ from $(\bU, \bW_{\text{Enc}})$:
\begin{equation}\label{eq:encoder}
\bW_{\text{Enc}}=\Psi(f)=\int_{\R^D} f(\bv)\bw(\bv) \diff \bv, \quad \Phi(\bU, \bW_{\text{Enc}})(\bv) = \exp(h(\bv; \bU, \bW_{\text{Enc}})).
\end{equation}
The exponential form is chosen in order to ensure positivity of the reconstructed density.
As an auxiliary goal, we would also like to predict a macroscopic analog of the entropy $\eta$ 
$$\eta(f)=\int_{\R^D}-f\ln{f}\diff\bv,$$
with all the available moments.
Accordingly, the objective function to be minimized reads 
\begin{equation}
\label{eq:auto_loss}
    {\E}_{f\sim \cD} [\|f - \Phi(
    \bU, \Psi(f))\|^2
    + \lambda_\eta (\eta(f) - h_\eta(\bU, \Psi(f)))^2].
\end{equation}
In~\eqref{eq:auto_loss}, the unknown functions to optimize are $\bw, h, h_\eta$ and the trial space (also known as the hypothesis space in machine learning)
for all three functions can be any machine learning models.  In this work we choose them to be 
multilayer feedforward neural networks. 
In practice the $f$'s are always discretized into finite dimensional vectors
and the quantities  in \eqref{eq:auto_loss} are actually squared $l^2$ norms of the associated vectors.
Once the optimal functions $\bw, h, h_\eta$ are trained, $\bW_{\text{Enc}}=\Psi(f)$, as a general alternative to $\bW_\text{Herm}$, provides a new set of the generalized moments of the system. 

\subsection{Learning Moment Closure}
\label{sec:learn_closure}
Recall the dynamic equation~\eqref{moment_dynamics} for the moment system, the goal of moment closure is to find suitable 
approximations of  $\bF, \bG, \bR$ as functions of $(\bU, \bW)$.
We first rewrite~\eqref{moment_dynamics} into
\begin{equation}
\label{moment_dynamics_v2}
  \left\{
  \begin{aligned}
     & \dt \bU + \dx\cdot [\bF_0(\bU) + \tF(\bU, \bW)] = 0, \\
     & \dt \bW + \dx\cdot [\bG_0(\bU) + \tG(\bU, \bW)] = \frac{1}{\veps}\bR(\bU, \bW),
  \end{aligned}
  \right.
\end{equation}
where $\bF_0(\bU), \bG_0(\bU)$ are the fluxes of the corresponding moments $\bU, \bW$ under the local Maxwellian distribution, 
that is, $\bF_0(\bU) \equiv \bF_{\text{Euler}}(\bU)$ and 
\begin{equation*}
    \bG_0(\bU) = \int_{\R^D}f_M(\bv; \bU) \bw(\bv)\bv\transpose \diff\bv.
\end{equation*}
Here we denote by $f_M(\cdot; \bU)$ the Maxwellian distribution defined by the macroscopic variables $\bU$. 
Our experience has been that separating the terms $\bF_0(\bU)$ and $\bG_0(\bU)$ out from $\bF$ and $\bG$ serves to reduce the variance during training. In practice, we can  calculate $\bw(\cdot)$ on a few points in advance and use the Gauss-Legendre quadrature rule to approximate $\bG_0(\bU)$ efficiently. 
Now our goal becomes finding  $\tF, \tG, \bR$.
Note that the first component of $\tF$, corresponding to the correction of the density flux, is always zero since the mass continuity equation is exact.

Take learning $\tG, \bR$ for example. At this point, we need a more concrete definition for the dynamics 
encoded in \eqref{moment_dynamics_v2}. The easiest way of doing this is to introduce certain
 numerical scheme for \eqref{moment_dynamics_v2}. We should emphasize that this scheme only serves to define
 the dynamics for the PDE. The machine learning model obtained is independent of the details of this scheme so long as it
 defines the same dynamics (in the limit as the grid size goes to 0).
We can think of a numerical scheme $\mathcal{S}$ as an operator, which takes the flux function, the stencil, and numerical discretization parameters as input, and outputs the increment of the function values at the targeted point in time. 
For instance,  in the case when $D=1$, a simple example of $\mathcal{S}$  may take the form
$$\hat{\bW}_{j,n+1} - \bW_{j,n} \approx \mathcal{S}[\tG,\bR](\bU_{j-1,n}, \bU_{j,n}, \bU_{j+1,n}, \bW_{j-1,n}, \bW_{j,n}, \bW_{j+1,n}; \Delta t, \Delta x, \veps)$$
where $j$ and $n$ denote the spatial and temporal indices  respectively.
In this case the tuple $\mathcal{X} = (\bU_{j-1,n}, \bU_{j,n}, \bU_{j+1,n}, \bW_{j-1,n}, \bW_{j,n}, \bW_{j+1,n}, \bW_{j,n+1})$ is
an example of data needed. 
The loss function can be chosen as:
\begin{equation*}
    {\E}_{f\sim \cD} \| \bW_{j,n+1} -  \bW_{j,n} - \mathcal{S}(\bU_{j-1,n}, \bU_{j,n}, \bU_{j+1,n}, \bW_{j-1,n}, \bW_{j,n}, \bW_{j+1,n})\|^2.
\end{equation*}
Care should be exercised when choosing the numerical scheme, since the solutions of these systems may contain shocks.
The task for learning $\tF$ can be formulated similarly based on the dynamics of $\bU$. 
The details of the numerical scheme we use can be found in SI Appendix, Sec.~B.

\subsection{Data Exploration}
The quality of the proposed moment method depends on the quality of the data set $\cD$ that we use to train the model. 
It consists of several solutions of the original Boltzmann equation~\eqref{eq:Boltzmann} under different initial conditions.
Unlike most conventional machine learning problems that rely on fixed given data sets, here the construction of the data set is completely our own choice, and is an important part of the algorithm.
In general our objective is  to achieve greater accuracy with fewer training data by choosing the training data wisely. 
In this sense it is close to that of  active learning~\cite{settles2009active}.
To achieve this, an interactive algorithm is required between the augmentation of the data set and the learning process.

In this work we adopt the following strategy. 
One starts with a relatively small data set and uses it
to learn the models. 
Then a new batch of solutions are generated for both the original kinetic model~\eqref{eq:Boltzmann} and the moment system~\eqref{moment_dynamics_v2}. The error in
the macroscopic variables $\bU$  is calculated as an indicator and the ones with large errors are added to the data set for the next round of learning.
These two steps are repeated  until convergence is achieved, which indicates that the phase space has been sufficiently explored. 
The whole scheme works as a closed loop and forms a self-learning process.

One key question is how to initialize the new batch of solutions. In principle we would like to initialize them
so that at the end of the active learning process, the configurations that occur in practice have been explored sufficiently.
Unfortunately at the moment, there are no precise mathematical principles that we can use to quantify this.
This is certainly one issue that we will continue to investigate in the future.
More details of the exploration procedure used can be found in SI Appendix, Sec.~E.

\subsection{Symmetries and Galilean Invariant Moments}
\label{sec:galilean}
An important issue in building machine learning models is how to handle the symmetries
in the system. The handling of translational, rotational and even permutational symmetries has been discussed in depth
in the literature already~\cite{ling2016reynolds,zaheer2017deep,zhang2018end,eickenberg2018solid}.
Besides these static symmetries,  Boltzmann equation also possesses an important  dynamic symmetry, the Galilean invariance.
Specifically, for every $\bu'\in \R^D$, define $f'$ by  $f'(\bx, \bu, t) = f(\bx-t\bu', \bv-\bu', t)$. If $f$ is a solution of the Boltzmann equation, 
then so is $f'$. It is desirable that the reduced moment system also inherits such an invariance. 
Here we present a viable approach that achieves this goal.

The idea is to define the generalized moments $\bW_\Gal$ (the subscript $\Gal$ signifies  Galilean invariance) properly such that they are Galilean invariant. Given the velocity $\bu$ and  temperature $T$ of $f$, we modify the encoder to be
\begin{equation}
\label{eq:gal_encoder}
    \bW_\Gal=\Psi_\Gal(f)=\int_{\R^D} f(\bv) \bw\left(\frac{\bv-\bu}{\sqrt{T}}\right) \diff \bv.
\end{equation}
Note that the encoder $\Psi_\Gal$ now depends nonlinearly on the first and second moments of $f$ 
(through $\bu$ and $T$) and is invariant with respect
to the choice of the Galilean reference frame.
It is straightforward to see that the Grad-type moments $\bW_{\text{Herm}}$ is a special case of~\eqref{eq:gal_encoder} and thus are Galilean invariant.
Modeling the dynamics of $\bW_\Gal$ becomes more subtle due to the spatial dependence in $\bu, T$.
Below for simplicity we will work with a discretized form of the dynamic model.
Suppose we want to model the dynamics of $\bW_{\Gal,j}$ at the spatial grid point indexed by $j$. Integrating the Boltzmann equation 
against  the generalized basis at this grid point gives
\begin{equation}
\label{eq:galilean1}
    \partial_t \int_{\R^D}f\bw\left(\frac{\bv-\bu_j}{\sqrt{T_j}}\right)\diff\bv + \dx \cdot \int_{\R^D}f\bw\left(\frac{\bv-\bu_j}{\sqrt{T_j}}\right)\bv\transpose \diff \bv = \int_{\R^D}\frac1\veps Q(f)\bw\left(\frac{\bv-\bu_j}{\sqrt{T_j}}\right)\diff \bv.
\end{equation}
The collision term on the right-hand side evaluated at the grid point $j$ can still be approximated reasonably well by a function of $(\bU_j, \bW_{\Gal,j})$ only since there is no spatial interaction involved. 
However, after the discretization, the flux term above is going to depend not only on $(\bU, \bW_\Gal)$ but also on the basis quantities $(\bu_j, T_j)$ chosen in~\eqref{eq:galilean1}. 
This motivates us to consider the following approximate moment equation
\begin{equation}
\label{eq:galilean_dynamics}
    \partial_t \bW_\Gal + \dx \cdot \bG_\Gal(\bU, \bW_\Gal; \bU_j) = \frac{1}{\veps}\bR_\Gal(\bU, \bW_\Gal).
\end{equation}
Note that  \eqref{eq:galilean_dynamics} is only meant to be used to model the dynamics of $\bW_{\Gal,j}$.
To model the dynamics of $\bW_{\Gal,j'}$ at another grid point $j'$, a different basis information $\bU_{j'}$ is provided.
Given the moment equation~\eqref{eq:galilean_dynamics}, the loss functions for $\bG_\Gal, \bR_\Gal$ can be defined in  the same way as in Sec.~\secclosure.
More discussion on the dynamics of Galilean invariant moment systems can be found in SI Appendix, Sec.~D.

One should also note that while preserving invariances is an important issue,
it is not absolutely necessary to preserve such invariances exactly.
If we make the trial space too restrictive in order to preserve invariances,  it may become difficult to find a model with satisfactory accuracy.
On the other hand,  if the reduced dynamics is sufficiently accurate,
all the invariances of the original system should be satisfied approximately with similar accuracy.

\subsection{An End-To-End Learning Procedure}
\label{sec:e2e}
In the previous sections we have introduced a two-step procedure to learn separately the moments $\bW_{\text{Enc}}$ (or $\bW_{\text{Gal}}$) and the dynamics of the moments. Here we also present an alternative end-to-end learning procedure.

For simplicity, we go back to the framework of Sec.~\secenc and Sec.~\secclosure
(i.e. using $\bW_{\text{Enc}}$ without specifying explicitly the Galilean invariance) in the 1-D case. 
The loss function for the autoencoder component is still defined as in ~\eqref{eq:auto_loss}.
The functions in the moment equations are learned using the following loss functions,
guided by \eqref{eq:integral_Boltz} and its approximation~\eqref{moment_dynamics_v2}. 
For $\tilde{\bF}$, only the third component $\tilde{\bF}_3$, the energy flux, needs to be 
considered because the mass and momentum continuity equations in \eqref{euler_full} are exact in the 1-D case. 
The corresponding loss function is given by
\begin{equation}
\label{eq:e2e_loss2}
    {\E}_{f\sim \cD} \|\int_{\R}\frac12f|\bv|^2\bv\transpose \diff \bv - (E+p)\bu\transpose - \tF_3(\bU,\bW_{\text{Enc}})\|^2.
\end{equation}
Similarly, the loss function for learning $\tG$ and $\bR$ can be written as
\begin{align}
    &{\E}_{f\sim \cD} \|\int_{\R}f\bw(\bv)\bv\transpose \diff \bv - \int_{\R}f_M(\bv;\bU)\bw(\bv)\bv\transpose \diff \bv - \tG(\bU,\bW_{\text{Enc}})\|^2, \label{eq:e2e_loss3}\\
    &{\E}_{f\sim \cD} \|\int_{\R} Q(f)\bw(\bv) \diff \bv - \bR(\bU,\bW_{\text{Enc}})\|^2. \label{eq:e2e_loss4}
\end{align}
Linearly combining eqs.~\eqref{eq:auto_loss} and \eqref{eq:e2e_loss2} to \eqref{eq:e2e_loss4} 
defines a loss function that allows us to learn the moment system in  a single optimization step. The benefits of such a learning strategy are twofold. 
On one hand, any single snapshot $f_n$ ($n$ denotes the temporal index when solving the kinetic equation)
is sufficient for evaluating the loss function in the end-to-end approach while two consecutive discretized snapshots $f_n, f_{n+1}$ 
are required in the loss function described in Sec.~\secclosure and Sec.~\secgalilean.
The resulting paradigm becomes more like unsupervised learning rather than supervised learning since solving the Boltzmann equation becomes unnecessary. On the other hand,  accuracy can potentially be improved since all the parameters are optimized jointly.

\subsection{An Alternative Direct Machine Learning Strategy}
\label{sec:directconv}
A more straightforward machine learning approach is to stay at the level of the original hydrodynamic variables $\bU$ and directly learn a correction term for the Euler equations to approximate the dynamics of the Boltzmann equation:
\begin{equation}
    \dt \bU +\dx\cdot(\bF_{\text{Euler}}(\bU) + \tF [\bU (\cdot);\veps])=0.
\end{equation}
Here $\tF [\bU ( \cdot);\veps]$ is a functional of $\bU(\cdot)$ that depends on  $\bU$, where $\bU$ is viewed as a function on
 the whole space. 
It is quite straightforward to design machine learning models under this framework.
For instance, assume that $D=1$. One can discretize  $\bU$ using $N_x$ grid points and train a network to approximate
$\tF$ in which the input is an $N_x\times 4$ matrix representing $\bU$ and $\veps$, and the output are the values of $\tF$ at the $N_x$ grid points.
In practice, to guarantee translational  invariance and approximate locality, we choose a $1$-D convolutional neural network with small convolutional kernels. 
This approach is conceptually simple and easy to implement.
Ideas like this have been used in various problems, including the turbulence models~\cite{ling2016reynolds,duraisamy2019turbulence,fonda2019deep}.

There are several problems with this. The first is that the approach does not offer any room for improving accuracy
since no  information can be used other than the instantaneous hydrodynamic variables.
The second is that this procedure requires a specific discretization to begin with.  The model
obtained is tied with this discretization. We would like to have machine learning models that are more like  PDEs.
The third is that the models obtained is harder to interpret. For instance the role that the Knudsen number plays in the convolutional network is not as explicit as in the Boltzmann equation or the moment equation.
In any case, it is our views that the moment system is a more appealing approach.

\section{Numerical Results}
\label{sec:numerical}
In this section we report results for the methods introduced above for 
the BGK model \eqref{eq:Q_bgk} with $D=1$ 
and the Boltzmann equation for Maxwell molecules \eqref{eq:Q_maxwell} with $D=2$.
At the moment, there are no reliable conventional moment systems for the full Boltzmann equation for Maxwell molecules.
In contrast, the methodology introduced here does not make use of the specific form of $Q(f)$ and can be 
readily applied to this or even more complicated kinetic models.

We consider the 1-D interval $[-0.5, 0.5]$ in the physical domain with periodic boundary condition and the time interval $[0, 0.1]$. Note that for the 2-D Maxwell model, we assume the distribution function is constant in the $y$ direction of the physical domain so that it is still sufficient to compute the solution on the 1-D spatial interval.
Three different tasks are considered, termed \taska, \taskb, and \taskc respectively. 
In the first two tasks the Knudsen number $\veps$ is constant across the whole domain.
This constant value is sampled from a log-uniform distribution on $[-3, 1]$ respect to base 10, i.e., $\veps$ takes values from $10^{-3}$ to 10. For data exploration, the initial conditions used in \taska consist of a few combination of waves randomly sampled from a probability distribution. 
For \taskb the initial conditions contain a mixture of waves and shocks, both randomly sampled from their respective
probability distributions. More details can be found in SI Appendix, Sec.~A. 
In \taskc, the initial conditions are the same as the ones in \taskb but the Knudsen  number is no longer constant in the spatial domain. Instead it varies from $10^{-3}$ to 10.
This is a toy model for transitional flows. We do not train any new models in \taskc but instead adopt the model learned from \taskb 
 to check its transferability. 

We generate the training data with Implicit-Explicit (IMEX) scheme~\cite{filbet2010class} for the 1-D BGK model or fast spectral method~\cite{hu2012fast,gamba2017fast,hu2019fast} for the 2-D Maxwell model. The spatial and temporal domains are both discretized into 100 grid points. 
The velocity domain is discretized into 60 grid points in the 1-D BGK model and 48 grid points in each dimension in the 2-D Maxwell 
model.
In all the numerical experiments,
\taska uses 100 paths as the training data set whereas \taskb use 200 paths. 
All the results reported are based on 100 testing initial profiles sampled from the same distribution of the corresponding tasks.
To evaluate the accuracy on the testing data, we consider two types of error measures for the macro quantities $\bU$, the relative absolute error (RAE) and the relative squared error (RSE).
More details of the tasks and data are provided in SI Appendix, Sec.~A.

\begin{figure}[ht]
    \centering
    \includegraphics[width=0.98\textwidth]{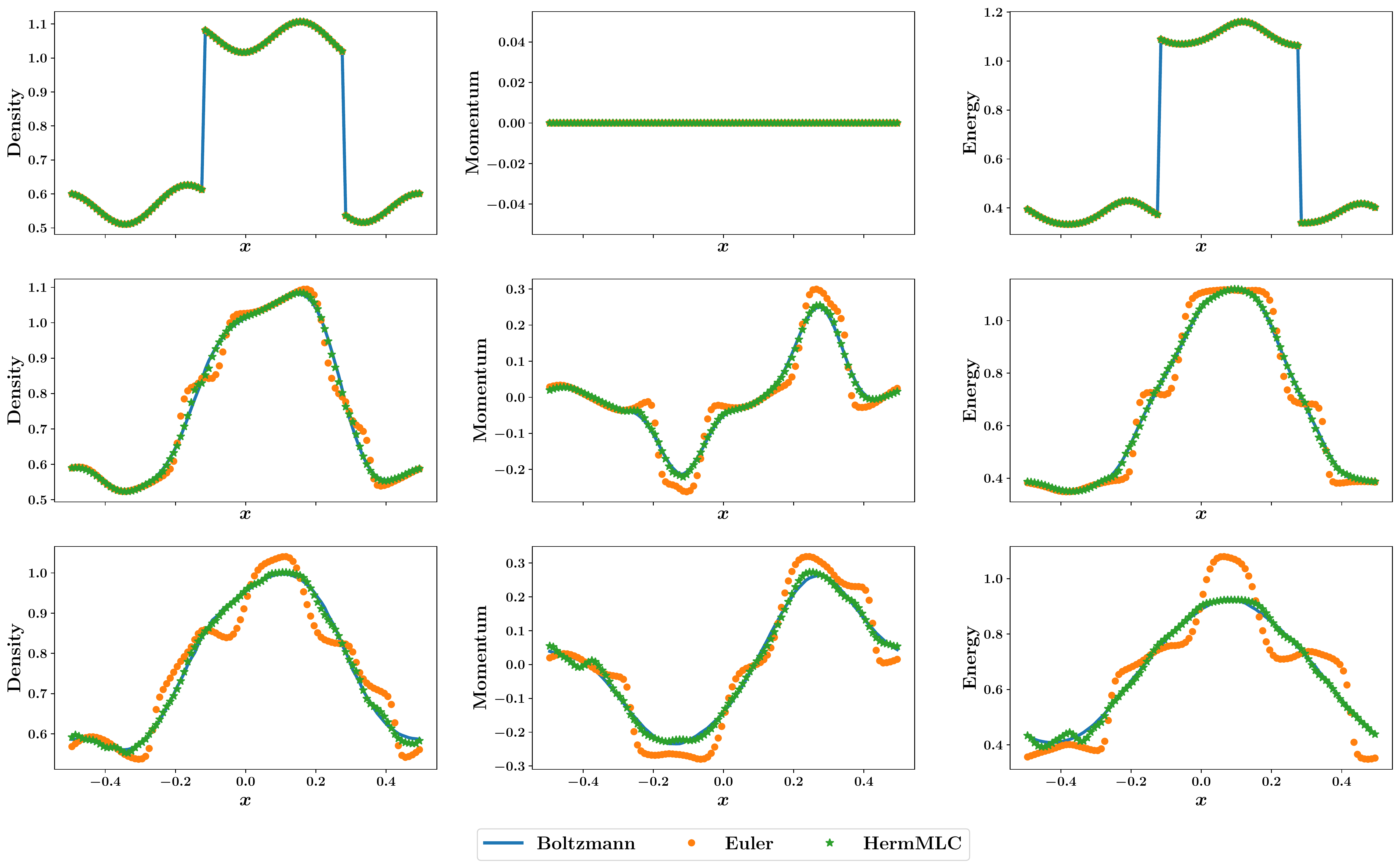}
    \caption{Sample profiles of mass, momentum, and energy densities (from left to right) at $t=0, 0.05, 0.1$ (from top to bottom) in  \taskb for the 2-D Maxwell model and $\veps=5.78$, obtained from the kinetic equation, the Euler equations, and HermMLC.}
    \label{fig:prof_mix_boltz_herm}
\end{figure}

We name different moment systems using a combination of the definition of $\bW$ and closure model. The moments considered in this paper, $\bW_{\text{Herm}}$, $\bW_{\text{Enc}}$, $\bW_{\text{Gal}}$ are abbreviated as Herm, Enc, GalEnc, respectively. The machine learning-based closure and its end-to-end version is called MLC and E2EMLC for short.
When $\bW_{\text{Enc}}$ is used, the learning of the closure takes the form introduced in Sec.~\secclosure and when $\bW_{\text{Herm}}$ or $\bW_{\text{Gal}}$ is used, it takes the form introduced in Sec.~\secgalilean.
The model described in Sec.~\secdirectconv is called DirectConv.
In all the numerical experiments presented in the main text, we always set $\bW_\text{Enc}$, $\bW_\text{Gal}$ to be 6-dimensional for the 1-D BGK model  and 9-dimensional for the 2-D Maxwell model. 
When Hermite polynomials are used for the moments, due to the numerical instability of evaluating moments of high-order polynomials, we limit the  order to be no larger than 5 for each variable. The resulting $\bW_{\text{Herm}}$ is 3-dimensional for the 1-D BGK model and 6-dimensional for the 2-D Maxwell model, dropping the ones that are identically $0$ due to symmetry.
Results of EncMLC are always augmented by data exploration, unless specified.
Other models do not use data exploration for ease of comparison.
Table~\ref{tab1} and Table~\ref{tab2} report the relative error of the different models on the testing data for the three tasks.
Fig.~\ref{fig:prof_mix_boltz_herm} shows an example of the profiles of the mass, momentum, and energy densities at $t=0, 0.05, 0.1$ for the same initial condition in \taskb for the 2-D Maxwell model, obtained by solving the kinetic equation, the Euler equations, and HermMLC. 

The benefits brought by data exploration is clearly shown through the comparison between the first two rows in Table~\ref{tab1}. EncMLC with data exploration has a similar accuracy compared to GalEncMLC. However, when the two models are trained on the data sets of the same size and distribution without data exploration, GalEncMLC performs better than EncMLC, for both the 1-D BGK model and 2-D Maxwell model.
The superiority of GalEncMLC on data efficiency is not surprising since it better captures the intrinsic features of the original dynamical system.

\begin{table}[ht]
\begin{tabular}{@{}l|ccc@{}}
\toprule
\thead[c]{\normalsize{Model}}         & \textit{Wave}          & \textit{Mix}     & \textit{MixInTransition}  \\ \midrule
EncMLC (no explor)         & 1.27(13), 1.60(35) & 1.68(10), 2.35(12)    & 1.82(11), 2.49(11) \\ 
EncMLC (explor)         & 0.85(10), 1.01(14) & 1.25(5), 1.75(8)    & 1.55(4), 2.06(7) \\ 
GalEncMLC & 0.97(22), 1.11(23) & 1.28(6), 1.85(10)    & 1.53(4), 2.11(8)\\
EncE2EMLC      & 0.76(6), 0.92(7) & --   & --\\
HermMLC     & 0.34(5), 0.43(6)   & 0.92(4), 1.45(10)    & 1.77(25), 3.01(27) \\
DirectConv       & 0.92(3), 1.16(6)   & 0.83(5), 1.13(9)     & 2.57(20), 3.61(26)\\
\bottomrule
\end{tabular}
\bigskip
\caption{Relative error (in percentages) of  the different machine learning models for the three tasks with the  BGK model. 
The first one in each cell denotes the relative absolute error (RAE) and the second  denotes the relative squared error (RSE). The numbers in the parentheses denote the standard deviation of the last one or two digits computed from three independent runs.
The results in the fourth column are obtained using the models learned in \taskb directly.}
\label{tab1}
\end{table}

\begin{table}[ht]
\begin{tabular}{@{}l|ccc@{}}
\toprule
\thead[c]{\normalsize{Model}}         & \textit{Wave}          & \textit{Mix}     & \textit{MixInTransition}  \\ \midrule
EncMLC (no explor)         & 0.83(7), 1.19(13) & 1.30(11), 1.96(16)    & 1.50(7), 2.28(10) \\ 
GalEncMLC & 0.77(6), 1.09(10) & 1.18(10), 1.77(15)    & 1.42(7), 2.12(13)\\
HermMLC       & 0.30(4), 0.50(10)   & 1.12(7), 1.74(11)     & 1.22(4), 1.89(5)\\
DirectConv       & 1.49(8), 2.22(14)   & 1.00(6), 1.47(9)     & 2.62(52), 3.64(74)\\ 
\bottomrule
\end{tabular}
\bigskip
\caption{Relative error (in percentages) of  the different machine learning models for the three tasks for the  Maxwell model. See Table~\ref{tab1} for more detailed explanation.}
\label{tab2}
\end{table}

\begin{sloppypar}
While the generalized moments and moment equations are learned differently in \taska and \taskb under the associated data distribution, in \taskc (the third column in Table~\ref{tab1} and Table~\ref{tab2}), we use the same models learned from \taskb directly without any further training.  The fact that the  relative error is similar to that in \taskb indicates that the machine learning-based moment system has satisfactory transferability.
On the contrary, while the model learned in \taska is accurate enough
for that particular task, it is not  sufficiently transferrable in general.
\end{sloppypar}

EncE2EMLC has the best accuracy in \taska for the BGK model among all the models based on $\bW_{\text{Enc}}$. Note that all the networks in EncE2EMLC have the same structure as in EncMLC, it seems that this improvement comes mainly from the end-to-end training process. 
However, when shocks are present, this model performs badly.
For example, it may produce unphysical solutions. This should be due to the lack of any enforcement of the entropy condition, either 
explicitly through the entropy function or implicitly through supervision from the dynamics of the kinetic equation.
On the other hand, the existence of shocks is a special feature of the physical problem considered here.
This issue disappears in most other physical systems and EncE2EMLC should become a very attractive approach for those
systems.

\begin{sloppypar}
The good performance of HermMLC suggests that the proposed machine learning-based closure is applicable to different types of moments.
The remarkable accuracy of HermMLC in \taska might be a result of
the close proximity between $f$ and the local equilibrium in this task. 
The fact that GalEncMLC achieved a similar accuracy
in \taskb and \taskc suggests that the autoencoder is an effective tool for 
finding generalized moments in general situations.
\end{sloppypar}

\begin{figure}[ht]
\centering
\includegraphics[width=0.98\textwidth]{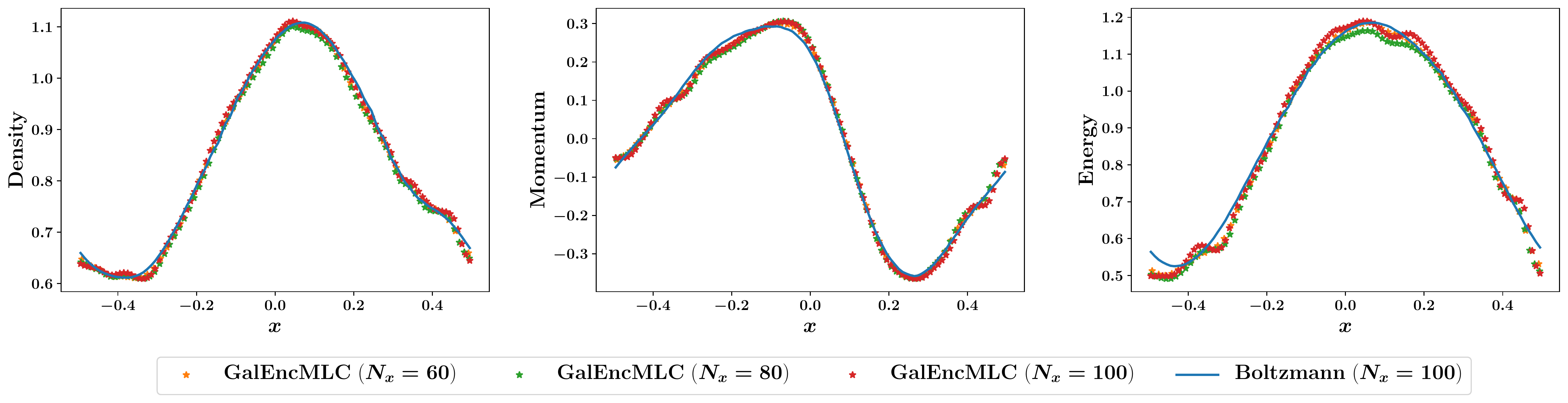}
\caption{The solutions of GalEncMLC at $t=0.1$ with the same initial profile as in \taskb but different spatial discretization, for the
2-D Maxwell model with $\veps=8.53$.}
\label{fig:nx_mix_boltz}
\end{figure}

The accuracy of DirectConv is quite good for both \taska and \taskb.
However, the model obtained is tied to the specific discretization
algorithm used and it is unclear how it can be used for other discretization schemes. 
The machine learning-based moment systems do not have this problem since they behave more like conventional PDEs. Fig.~\ref{fig:nx_mix_boltz} illustrates the solutions of GalEncMLC  under the same initial condition as in  \taska for the 2-D Maxwell model but different spatial discretization.

\begin{figure}[ht]
\centering
\includegraphics[width=0.96\textwidth]{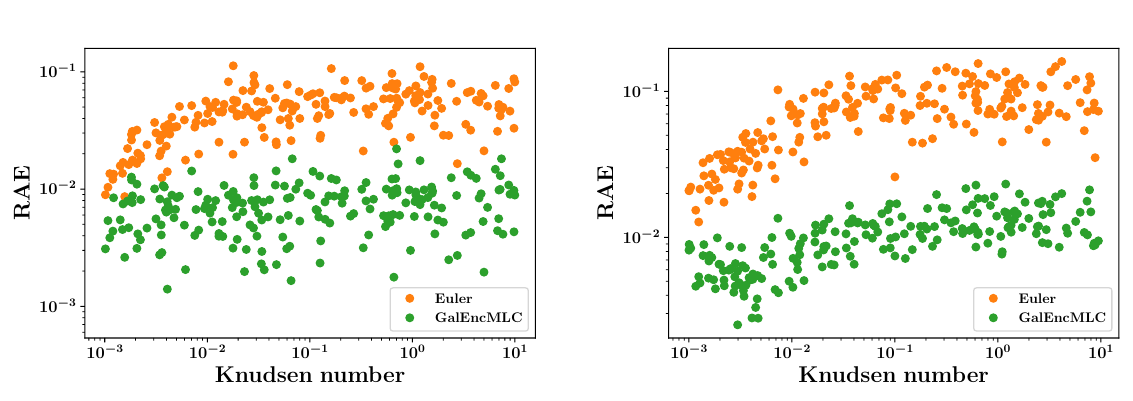}
\caption{The log-log scatter plot of the relative  error versus the Knudsen number $\veps$ for 200 testing paths in \taska (left) and \taskb (right) for the 2-D Maxwell model.}
\label{fig:scatter_boltz}
\end{figure}

In Fig.~\ref{fig:scatter_boltz} we display the log-log scatter plots of the relative  error versus the Knudsen number $\veps$ for both \taska and \taskb for the 2-D Maxwell model. 
One can see that the accuracy of the machine learning-based moment system is almost uniform across the whole regime, with the same computational cost. This stands in striking contrast to the conventional hydrodynamic models or DSMC method. 
As for the computational cost, to solve for one path for the 2-D Maxwell 
model on 
a Macbook Pro with a 2.9GHz Intel Core i5 processor, it takes about 5 minutes to run the fast spectral method for the original Boltzmann equation. In contrast, the runtime of the machine learning-based moment method (GalEncMLC) is only half a second.

We refer the interested readers to SI Appendix, Sec.~F for more numerical results, including the shape of the generalized basis functions in the 1-D BGK model, the accuracy under different number of moments, the solution on the test data under different spatial and temporal discretization, the growth of the relative error as a function of time in the three tasks, and more sample profiles in the three tasks.
A brief introduction of the hyperbolic regularized Grad's moment system~\cite{cai2014globally,cai2015framework}, a conventional moment method for the BGK model, and related results are provided in SI Appendix, Sec.~C.

\section{Discussion and Conclusion}
\begin{sloppypar}
This paper presents a new framework for multiscale modeling using machine learning in the absence of scale separation. 
We have put our emphasis on learning physical models, not just a particular algorithm.
We have studied some of the main issues involved, including the importance of obeying physical constraints, actively learning,
end-to-end models, etc. Our experience suggests that it is often advantageous to respect physical constraints, but there is no
need to sacrifice a lot of accuracy just to enforce them, since if we can model the dynamics accurately, the physical
constraints are also satisfied with similar accuracy.
Even though we still lack a proper mathematical framework to serve as guidelines, active learning is very important
in order to ensure the validity of the model under different physical conditions.
Regarding the end-to-end model, even though  it did not perform satisfactorily  for problem with shocks,
we feel that it may very well be the most  attractive approach in the more general cases since it seems 
most promising to derive uniform error bounds in this case.
\end{sloppypar}

There are a few important issues that need to be addressed when applying the methodology presented here to
other problems. The first is the construction of new relevant variables. Here our requirement is encoded in the autoencoder:
the reduced variables need to be able to capture enough information so that 
 the one-particle phase space distribution function can be accurately reproduced. 
 This needs to be generalized when considering other models,
for example  when the microscopic model is molecular dynamics.
The second is the starting point for performing ``closure''.  This component is also problem-specific.
Despite the fact that these important components need to be worked out for each specific problem, we do feel
that general procedure here is applicable for a wide variety of multiscale problems.

Going back to the specific example we studied,  the BGK model or the Boltzmann equation for Maxwell molecules, 
we presented an interpretable generalized moment system that works well over a wide range of Knudsen numbers.
One can think of these models just like conventional PDEs, except that some of the terms in the fluxes and forcing 
are stored as subroutines. This is not very different from the conventional Euler equations for complex gases where the
equations of state are stored as look-up tables or sometimes subroutines.
Regarding the three ingredients involved in learning the reduced models, namely, labeling the data, learning from the data, and exploring the data (as shown in Fig.~\ref{fig:schematic}),   labeling the data is straightforward in this case: we just need to solve the 
kinetic  equation for some short period of time under different initial conditions. Data exploration is carried out using Monte Carlo sampling from some prescribed initial velocity distributions, and the picking of these initial velocity distributions is still somewhat ad hoc.
The learning problem is the part that we have studied most carefully.  We have explored and compared
several different versions of  machine learning models. We are now ready to attack other problems such as 
kinetic models in plasma physics and kinetic models for complex fluids, building on this experience.

\section*{Acknowledgement}
The work presented here is supported in part by a gift to Princeton University from iFlytek and the 
Office of Naval Research grant N00014-13-1-0338.
We also thank the reviewers for the valuable comments which helped to improve the
quality of the work and the paper.

\bibliography{ref}{}
\bibliographystyle{unsrt}

\newpage
\appendix
\section*{Supplementary Information}
\renewcommand\thesubsection{\Alph{subsection}}
\subsection{Tasks and Data}
\label{secSI:data}
We consider  1-D interval $[-0.5, 0.5]$ in the physical domain with periodic boundary condition.  We put down a uniform grid of  $100$ grid points. 
Note that for the 2-D Maxwell model, we assume that the spatial domain is homogeneous along the $y-$axis so that $f$ and all the macro variables only depend on the $x$ coordinate.
The time interval considered is $[0, 0.1]$ with time step size 0.001. In the 1-D BGK model, the  velocity domain is truncated to $[-10, 10]$ and discretized using 60 nodes according to the Gauss-Legendre quadrature rule. 
In the 2-D Maxwell model, the  velocity domain is truncated to $[-7.7, 7.7]\times [-7.7, 7.7]$ and discretized 
using 48 equidistant nodes in each dimension. 
In  \taska and \taskb the Knudsen number $\veps$ is sampled from a log-uniform distribution on $[-3, 1]$ respect  base 10, i.e., $\veps$ takes values from $10^{-3}$ to 10, constant across the domain. 
We consider two types of initial conditions in all the three tasks. 

For \taska the initial condition $f_{wave}$  is sampled from a mixture of two local Maxwellian distributions. Two macroscopic functions $\bU_1$, $\bU_2$ are  sampled from sine waves
\begin{equation*}
    \begin{cases}
    \rho(x,0) = a_\rho\sin(2k_\rho\pi x/L + \psi_\rho) + b_\rho, \\
    u(x,0) = 0, \\
    T(x,0) = a_T\sin(2k_T\pi x/L + \psi_T) + b_T.
    \end{cases}
\end{equation*}
Here we assume $a_\rho, \psi_\rho, b_\rho$ are random variables sampled from the uniform distributions on $[0.2, 0.3]$, $[0, 2\pi]$, $[0.5, 0.7]$, respectively. $k_\rho$ is a random integer sampled uniformly from the set $\{1, 2, 3, 4\}$. $a_T, \psi_T, b_T, k_T$ in  $T(\cdot,0)$ are independent and identically distributed random variables as their counterparts in the function $\rho(\cdot,0)$. Finally, two local Maxwellian distributions are randomly mixed through
$$f_{wave} = \frac{\alpha_1 f_M(\bv;\bU_1) + \alpha_2 f_M(\bv;\bU_2)}{\alpha_1 + \alpha_2 + 10^{-6}},$$
in which $\alpha_1, \alpha_2$ are two random variables sampled from the uniform distribution on $[0, 1]$.

For \taskb the  initial condition $f_{mix}$  is sampled from a random superposition of two functions, $f_{wave}$ as defined above and $f_{shock}$.  $f_{shock}$ is also point-wise local Maxwellian, except that the macroscopic  functions $\bU$ are made up from
some  Riemann problems.
Consider the Riemann problem in which $\rho_L, T_L$ are two independent random variables sampled from the uniform distribution on $[1, 2]$, $\rho_R, T_R$ are two independent random variables sampled from the uniform distributions on $[0.55, 0.9]$, and $u_L, u_R$ are 0. The initial condition $f_{shock}$ then has the form
\begin{align*}
    &\begin{cases}
        \rho(x,0), u(x,0), T(x,0) = \rho_L, u_L, T_L, &x\in[-0.5,x_1] \text{ or } x\in[x_2, 0.5],  \\
        \rho(x,0), u(x,0), T(x,0) = \rho_R, u_R, T_R, &x\in(x_1, x_2),
    \end{cases}\\
    \text{or } &\begin{cases}
        \rho(x,0), u(x,0), T(x,0) = \rho_R, u_R, T_R, &x\in[-0.5,x_1] \text{ or } x\in[x_2, 0.5],  \\
        \rho(x,0), u(x,0), T(x,0) = \rho_L, u_L, T_L, &x\in(x_1, x_2),
    \end{cases}
\end{align*}
where $x_1, x_2$ are two random variables sampled from the uniform distributions on $[-0.3,-0.1]$ and $[0.1, 0.3]$.
Finally, we linearly combine two initial conditions to obtain
$$f_{mix}=\alpha f_{wave} + (1-\alpha) f_{shock},$$ where $\alpha$ is a random variable sampled from the uniform distribution on $[0.2, 0.6]$.

\begin{figure}[ht]
\centering
\includegraphics[width=0.98\textwidth]{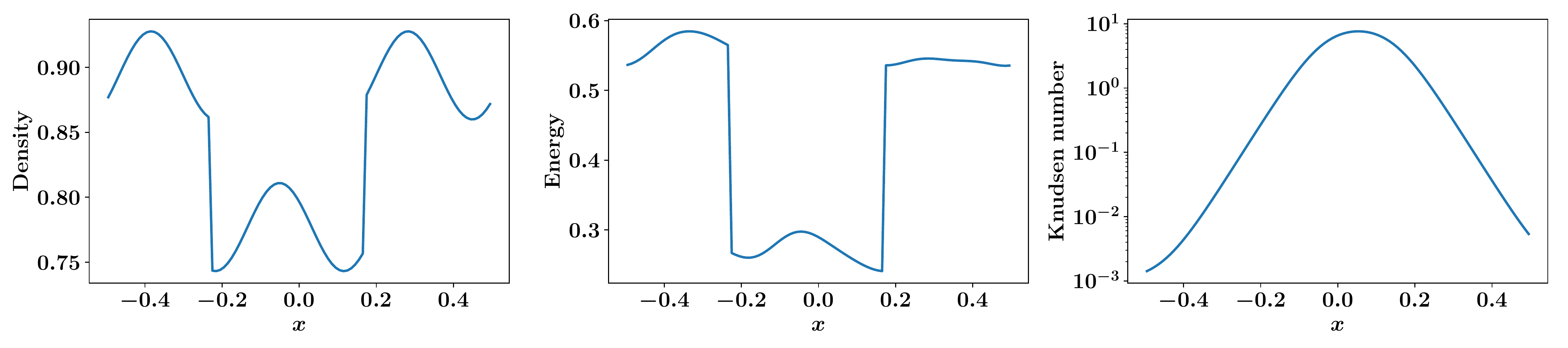}
\caption{Sample profiles of mass and energy densities, and Knudsen number in  \taskc.}
\label{fig:kn_profile}
\end{figure}

For \taskc the initial condition  is the same as in \taskb.  The values of $\veps$ vary from  $10^{-3}$ to 10 in the domain,
  similar to the one used in~\cite{FILBET20107625},
$$\veps(x)=10^{-3} + 5(\tanh(1+11(x-x_0))+\tanh(1-11(x-x_0))),$$
where $x_0$ is sampled from the uniform distribution on $[-0.2, 0.2]$.  This is where the  Knudsen number is at its maximum. The sample profiles of mass and energy densities, and Knudsen number in  \taskc are shown in Fig.~\ref{fig:kn_profile}.

We compute two types of error, the relative absolute error (RAE) and the relative squared error (RSE), to measure the accuracy of different models. By a slight abuse of notation, we consider $N$ independent profiles of conserved quantities, $\bU^{(i)}$, $\hat{\bU}^{(i)}$, $i=1,\dots, N$, computed from the kinetic equation and the  machine learning-based model respectively. Assume that each profile is discretized using $N_x$ grid points indexed by  $j$. We define
\begin{align*}
    &\text{RAE} = \frac{1}{2NN_x}\sum_{i,j}
    \frac{|\bU^{(i)}_j -\hat{\bU}^{(i)}_j|}{|\bU^{(i)}_j|},\\
    &\text{RSE} = \sqrt{
    \frac{\sum_{i,j}(\bU^{(i)}_j -\hat{\bU}^{(i)}_j)^2}{\sum_{i,j}(\bU^{(i)}_j)^2}}.
\end{align*}
There are other ways to measure the accuracy,   but our experience suggests that the overall behavior is quite independent of the accuracy measure that we use.

\subsection{Numerical Scheme}
\label{secSI:scheme}
Here we introduce in detail the numerical scheme $\mathcal{S}$ used when learning the moment closure.
This enters into the concrete form of the loss function for the dynamics of $\bU$ and $\bW$. Recall the dynamic equation
\begin{equation}
  \left\{
  \begin{aligned}
     & \dt \bU + \dx\cdot [\bF_{\text{Euler}}(\bU) + \tF(\bU, \bW)] = 0, \\
     & \dt \bW + \dx\cdot [\bG(\bU, \bW)] = \frac{1}{\veps}\bR(\bU, \bW),
  \end{aligned}
  \right.
  \label{moment_dynamics_v3}
\end{equation}
where $\bG(\bU) = \bG_0(\bU) + \tG(\bU, \bW)$, as explained in Sec.~\secclosure.
Applying the classical finite volume discretization to \eqref{moment_dynamics_v3}, we get
\begin{equation}
\label{eq:scheme1}
    \left\{
    \begin{aligned}
        &\hat{\bU}_{j, n+1} = \bU_{j, n} - \frac{\Delta t}{\Delta x}\left((\bF_{\text{Euler}})_{j+1/2, n} - (\bF_\text{Euler})_{j-1/2, n}\right) - \frac{\Delta t}{\Delta x}(\tF_{j+1/2, n} - \tF_{j-1/2, n}), \\
        &\hat{\bW}_{j, n+1} = \bW_{j, n} - \frac{\Delta t}{\Delta x}(\bG_{j+1/2, n} - \bG_{j-1/2, n}) + \frac{\Delta t}{\veps} \bR(\bU_{j,n}, \bW_{j, n}),
    \end{aligned}
    \right.
\end{equation}
where $\hat{\bU}_{j,n+1}$ and $\hat{\bW}_{j,n+1}$ denote the one-step solution of the machine learning-based model at the spatial grid point $j$ and time step $n+1$. This is a conservative scheme with first order accuracy in both time and space. 

For  $(\bF_{\text{Euler}})_{j+1/2, n}$, any classical numerical scheme for solving the Euler equations can be applied since there is no parameter to optimize. In our implementation, we choose the $1$-D HLLC Riemann solver~\cite{Toro1997} with entropy fix,
implemented in the  open source package Clawpack~\cite{clawpack, mandli2016clawpack, LeVeque-FVMHP}.

Regarding the fluxes $\tF_{j+1/2, n}$ and $\bG_{j+1/2, n}$, 
we choose a generalized Lax-Friedrichs scheme based on a three-point stencil,
\begin{align}
\renewcommand*{\arraystretch}{1.3}
\begin{pmatrix}
\tF_{j+1/2, n} - \tF_{j-1/2, n}\\
\bG_{j+1/2, n} - \bG_{j-1/2, n}
\end{pmatrix} 
= &
\renewcommand*{\arraystretch}{1.3}
\begin{pmatrix}
\tF(\bU_{j+1, n}, \bW_{j+1, n}) - \tF(\bU_{j-1, n}, \bW_{j-1, n})\\
\bG(\bU_{j+1, n}, \bW_{j+1, n}) - \bG(\bU_{j-1, n}, \bW_{j-1, n})
\end{pmatrix} \notag\\
&~-
\renewcommand*{\arraystretch}{1.3}
\begin{pmatrix}
A_{\bU}(\bU_{j+1, n} -2\bU_{j, n} + \bU_{j-1, n})\\
A_{\bW}(\bW_{j+1, n} -2\bW_{j, n} + \bW_{j-1, n})
\end{pmatrix}
,
\label{eq:scheme2}
\end{align}
where the constants $A_{\bU}, A_{\bW}$ are two diagonal matrices denoting the  numerical viscosity coefficients. We optimize these constants  during training such that a suitable strength of the numerical viscosity can be found for the machine-learned fluxes.

Combining~\eqref{eq:scheme1} and \eqref{eq:scheme2}, we see that the one-step output of the numerical scheme $\mathcal{S}$ is continuous respect to all the parameters. Hence we can use stochastic gradient descent to optimize them. 

\subsection{Hyperbolic Moment Method}
\label{secSI:hyperbolic_mom}
\begin{sloppypar}
In this section, we briefly introduce a well-acknowledged moment method in the literature suitable for solving the BGK model, the hyperbolic regularized Grad's moment system (see for example \cite{cai2014globally,cai2015framework}).
As introduced in Sec.~\secenc, the starting point is the Hermite expansion of $f$~\eqref{eq:f_expansion}. The considered moments are the expansion coefficients $f_{\bm\alpha}$ truncated at a predefined order $L$.  
Plugging the expansion~\eqref{eq:f_expansion} into the Boltzmann equation and noting that the BGK collision operator can also be expressed using the basis function, one can collect all the coefficients for each basis function and deduce a system of equations. The system is closed by setting $f_{\bm\alpha}=0$ for all the $|\bm \alpha|>L$.
Finally a regularization term is added to the equation at the highest order to ensure that the moment system is hyperbolic.
We refer the readers to~\cite{MR2729444, MR2886321, MR3002565, MR3070796} for more details of this approach and its numerical implementation.
\end{sloppypar}

\begin{sloppypar}
The machine learning-based approach shares a lot in common with this conventional approach: both attempt
to solve the kinetic equation accurately; the issues of Galilean invariance are similar.
However, there are two important differences. First, while the approach mentioned above can be viewed as a spectral method for the $\bv$ component of the variables, the machine learning-based generalized moments presented has more flexibility in representing the data with adaptive basis functions.
Second, while the approach mentioned above is only suitable for a few relaxation types of collision operator (the Maxwell model is not included), the machine learning-based closure presented is equally applicable to other more realistic collision models.
\end{sloppypar}
In order to get some ideas about quantitative comparison, we implemented the algorithm for solving the hyperbolic regularized moment equation according to~\cite{MR3070796} and tested it on all the three tasks considered in the paper for the 1-D BGK model. We use $200$ grid points in the spatial domain and $9$ moments in total. The relative RAE and RSE in percentages are 2.36(3), 3.07(4) (\taska), 2.13(2), 3.25(17) (\taskb), 2.11(7), 3.39(8) (\taskc).
This is slightly worse than the results obtained from the machine learning-based models.
It should be pointed out that it is difficult to compare these results directly with the proposed machine learning-based moment system since there are a lot of factors that contribute to the  accuracy.
Nevertheless,  we feel that for more challenging problems, the advantage of the machine learning-based approach will be more striking, not alone its versatility to other collision models.

\subsection{Galilean Invariant Dynamics}
\label{secSI:galilean}
\begin{sloppypar}
The first step of GalEncMLC, finding generalized moments $\bW_\Gal$, naturally obeys Galilean invariance. It is more subtle for
the dynamics of $\bW_\Gal$ to obey Galilean invariance since the associated PDE has additional convection terms involving $\nabla_{\bx}\bu, \nabla_{\bx} T$ compared to \eqref{moment_dynamics}. 
The simplest approximate solution to this problem is to introduce some spatial dependence into the PDE models.
\end{sloppypar}

As discussed in Sec.~\secgalilean, in consideration of the local dynamics of $\bW_{\Gal,j}$ at the grid point $j$, the exact flux function should be 
$$\int_{\R^D}f(\bv)\bw\left(\frac{\bv-\bu_j}{\sqrt{T_j}}\right)\bv\transpose \diff \bv.$$ 
We need to approximate this quantity in the form of $\bG_\Gal(\bU, \bW_\Gal; \bU_j)$ as proposed in~\eqref{eq:galilean_dynamics}, a function of  the macroscopic variables $\bU, \bW_\Gal$, with the basis information $\bU_j$ provided as well. 
Meanwhile, we would like to have an alternative expression similar to the flux terms in ~\eqref{moment_dynamics_v2}
in order to reduce the variance during training. Note that in this setting $\bG_0(\bU)$ is much more expensive to compute than its counterpart in EncMLC since the Gauss-Legendre quadrature rule now requires evaluating the generalized basis at different sets of the grid points for each single batch of data, due to the nonlinear dependence on  $\bu$ and $T$. Instead we consider another decomposition of the flux function
\begin{align*}
    & \int_{\R^D}  f\bw\left(\frac{\bv-\bu_j}{\sqrt{T_j}}\right)\bv\transpose\diff \bv \\
    =~ &\int_{\R^D}  f\bw\left(\frac{\bv-\bu_j}{\sqrt{T_j}}\right)(\bv-\bu)\transpose\diff \bv
    + \left(\int_{\R^D}f\bw\left(\frac{\bv-\bu_j}{\sqrt{T_j}}\right)\diff \bv\right)\bu\transpose \\
    \approx~ &\int_{\R^D}  f\bw\left(\frac{\bv-\bu_j}{\sqrt{T_j}}\right)(\bv-\bu)\transpose\diff \bv
    + \left(\int_{\R^D}f\bw\left(\frac{\bv-\bu}{\sqrt{T}}\right)\diff \bv\right)\bu\transpose \\
    =~ & \int_{\R^D} f\bw\left(\frac{\bv-\bu_j}{\sqrt{T_j}}\right)(\bv-\bu)\transpose\diff \bv + \bW_\Gal\bu\transpose.
\end{align*}
Here the approximation $\bu\approx\bu_j$ in the second term is made on the grounds that the flux above is used only  locally around the grid point $j$. 
If we also assume $\bu\approx\bu_j$ in the first term above, it will be equivalent to $\int_{\R^D}  f(\bv+\bu)\bw(\frac{\bv}{\sqrt{T}})\bv\transpose\diff \bv$.
Such an expression is consistent with our consideration of the Galilean invariance because it only depends on the shape of the phase density rather than the choice of the Galilean reference frame.
The above decomposition finally motivates us to write 
\begin{equation}
\label{eq:galilean_flux}
\bG_\Gal(\bU,\bW_\Gal;\bU_j)=\tilde{\bG}_\Gal(\bU,\bW_\Gal;\bU_j) + \bW_\Gal\bu\transpose,
\end{equation}
where $\tilde{\bG}_\Gal$ is the neural network to be optimized. 
This ensures that the dynamics of $\bW_\Gal$ also obeys Galilean invariance approximately.

\subsection{Learning of Neural Networks}
\label{secSI:hyperparameter}
\subsubsection*{\bf EncMLC, GalEncMLC, and HermMLC}
We use the same architecture for the neural networks used in EncMLC, GalEncMLC, and HermMLC.
The input is always the concatenation of all the variables listed.
For the autoencoder, the basis function $\bw$ of the encoder in~\eqref{eq:encoder} is represented by a fully-connected neural network with two hidden layers and $3M$ hidden nodes in each layer (recall $M$ denotes the dimension of the generalized moments $\bW_\text{Enc}$ or $\bW_\text{Gal}$). We use the same technique as in Batch Normalization~\cite{Ioffe2015BN} to normalize  the output within each data batch  to ensure zero mean and unit variance. It is observed that this operation improves the stability of training.
The function $h$ in the decoder is represented by another neural network with two hidden layers, whose widths are $2M$ and $M$, respectively. The activation function is chosen to be the softplus function. $\lambda_\eta$ in \eqref{eq:auto_loss} is chosen to be 0.01. The Adam optimizer~\cite{Kingma2015adam} is used with learning rate 0.001 and batch size 100 for training the autoencoder. Usually the autoencoder is trained for $60$-$120$ epochs, depending on the type of initial condition and the size of the data set.

For the moment closure in~\eqref{moment_dynamics_v2},
$\tG$ and $\bR$ are modeled by two neural networks both with $3$ hidden layers and $64$ hidden nodes in each layer. 
For $\tF$ in the 1-D BGK model, only the component corresponding to energy is non-zero and it is  modeled by a neural network with two hidden layers and $32$ hidden nodes per layer.
For $\tF$ in the 2-D Maxwell model, the components corresponding to momentum and energy should be both approximated. The former is modeled by a neural network with two hidden layers and $8$ hidden nodes per layer and the latter is modeled by a neural network with two hidden layers and $32$ hidden nodes per layer.
All the neural networks use the residual connection~\cite{he2016deep} in the hidden layers. Again softplus is used as the activation function. When building the loss function, we have two separate terms: one for $\tF$ and the other for $\tG$ and $\bR$, as described in Sec.~\secclosure. We use a weighted sum of the two terms as our loss function, with equal weights $100$. An Adam optimizer is employed to train the networks. The learning rate starts from $0.01$ and decays exponentially to $0.001$. Training takes $20$ epochs with batch size $256$.

\subsubsection*{\bf EncE2EMLC}
For EncE2EMLC, we use the exact same architecture for all the networks as in EncMLC and GalEncMLC in both the autoencoder and the moment closure. The single loss function is a linear combination of ~\eqref{eq:auto_loss}\eqref{eq:e2e_loss2}\eqref{eq:e2e_loss3}\eqref{eq:e2e_loss4} with weights 0.01/$SS_1$, 0.01/$SS_2$, 0.01/$SS_3$, 0.01/$SS_4$, respectively. $\lambda_\eta$ is set to $0$ in~\eqref{eq:auto_loss}. 
Here $SS_1,\dots,SS_4$ denote the total sum of squares in the associated regression problem
\begin{align*}
    &SS_1 = {\E}_{f\sim \cD} \|f\|^2, \\
    &SS_2 = {\E}_{f\sim \cD} \|\int_{\R}\frac12f|\bv|^2\bv\transpose \diff \bv - (E+p)\bu\transpose\|^2, \\
    &SS_3 = {\E}_{f\sim \cD} \|\int_{\R}f\bw(\bv)\bv\transpose \diff \bv - \int_{\R}f_M(\bv;\bU)\bw(\bv)\bv\transpose \diff \bv\|^2, \\
    &SS_4 = {\E}_{f\sim \cD} \|\int_{\R} Q(f)\bw(\bv) \diff \bv\|^2.
\end{align*}
Hence the goal of the optimization is  to minimize the four relative losses with equal weights 0.01. Noting that $SS_3, SS_4$ actually depend on the encoder, we use the statistics within the batch during training to approximate them. An Adam optimizer is used for 90 epochs with batch size 100. The learning rate is  constant during each 30-epoch periods, deceasing from 0.005 to 0.001, and then to 0.0005.

\subsubsection*{\bf DirectConv}
As described in Sec.~\secdirectconv, we use a $1$-D convolutional neural network to represent the non-zero components of functional $\tF$, corresponding to the momentum and energy fluxes. The input of the network contains $4$ channels of length $N_x$. The architecture of the network can be expressed
$$
\begin{array}{ll}
 &N_x\times4\xrightarrow{Conv} N_x\times40 \xrightarrow{Pooling} N_x/2\times40 \xrightarrow{Conv} N_x/2\times40 \\
\xrightarrow{Pooling}& N_x/4\times40 \xrightarrow[Res]{Conv} N_x/4\times40 \xrightarrow[Res]{Conv} N_x/4\times40 \xrightarrow{Conv} N_x/4\times3 \\
\xrightarrow{Deconv}& N_x/2\times3 \xrightarrow{Deconv} N_x\times1 ~(\text{or~} N_x\times2)\\
\end{array}
$$
where $Conv$ represents $1$-D convolution with kernel size $4$ and periodic padding followed by a softplus activation, $Pooling$ 
is done by max pooling, $Res$ means a residual connection in the layer, and $Deconv$ means a $1$-D deconvolution operation with kernel size $4$ and stride $2$. The network is trained by an Adam optimizer with batch size $50$ and learning rate exponentially decreasing  from $0.001$ to $0.0002$. Training is run for $5000$ epochs.

\subsubsection*{\bf Data Exploration}
For \taska, an autoencoder is initially trained for $120$ epochs with a data set containing $50$ paths, sampled from the distribution of the initial profiles. Then, $5$ loops of exploration is done as follows. In each loop, $100$ new paths are evaluated by both the original kinetic model and the moment system, in which $10$ paths with the largest errors are added to the data set, then the autoencoder is retrained for another $20$ epochs on the new data set. Finally the data set contains $100$ paths, the same as in the case without data exploration. For \taskb the autoencoder is initially trained for $90$ epochs on $100$ paths, before $5$ loops of data exploration. In each loop, $20$ paths with the largest errors from $200$ randomly sampled paths are added to the data set, and then the autoencoder is retrained for $15$ epochs. The final data set contains $200$ paths, the same as in the case without data exploration. The autoencoders, fluxes, and production terms used in \taskc are the same as \taskb.
No additional training was used.

It is worth mentioning that in the current implementation of data exploration, the cost of generating truthful micro-scale data is not directly reduced because it is needed in evaluating the prediction error of the new data in the exploration stage. 
There are various ways to fix this problem,  for instance, by using the variance from the predictions of an ensemble of  networks optimized independently as an indicator of the error in the exploration, as was done in \cite{zhang2019active}. 
The ideal scenario is that given a fixed budget for generating the data, the exploration procedure should provide training samples of the highest quality so that the best testing performance can be achieved.
This is left for future work.

\subsection{Additional Results}
\label{secSI:res}
We train GalEncMLC for \taskb for the 1-D BGK model using 3 or 9 generalized moments, the  resulted RAE and RSE  are 2.00(12), 3.03(30) and 1.27(5), 1.92(18) respectively. 
We train GalEncMLC for \taskb for the 2-D Maxwell model using 6 generalized moments, the  resulted RAE and RSE are 1.39(10), 2.27(16).
Comparing these results with Table~\ref{tab1} and Table~\ref{tab2}, we see that choosing 6 additional generalized moments in the 1-D BGK model and 9 additional generalized moments in the 2-D Maxwell model is a suitable trade-off between accuracy and efficiency given the  sizes of the networks and the data sets used in the current experiments. 

\begin{figure}[ht]
    \centering
    \includegraphics[width=0.98\textwidth]{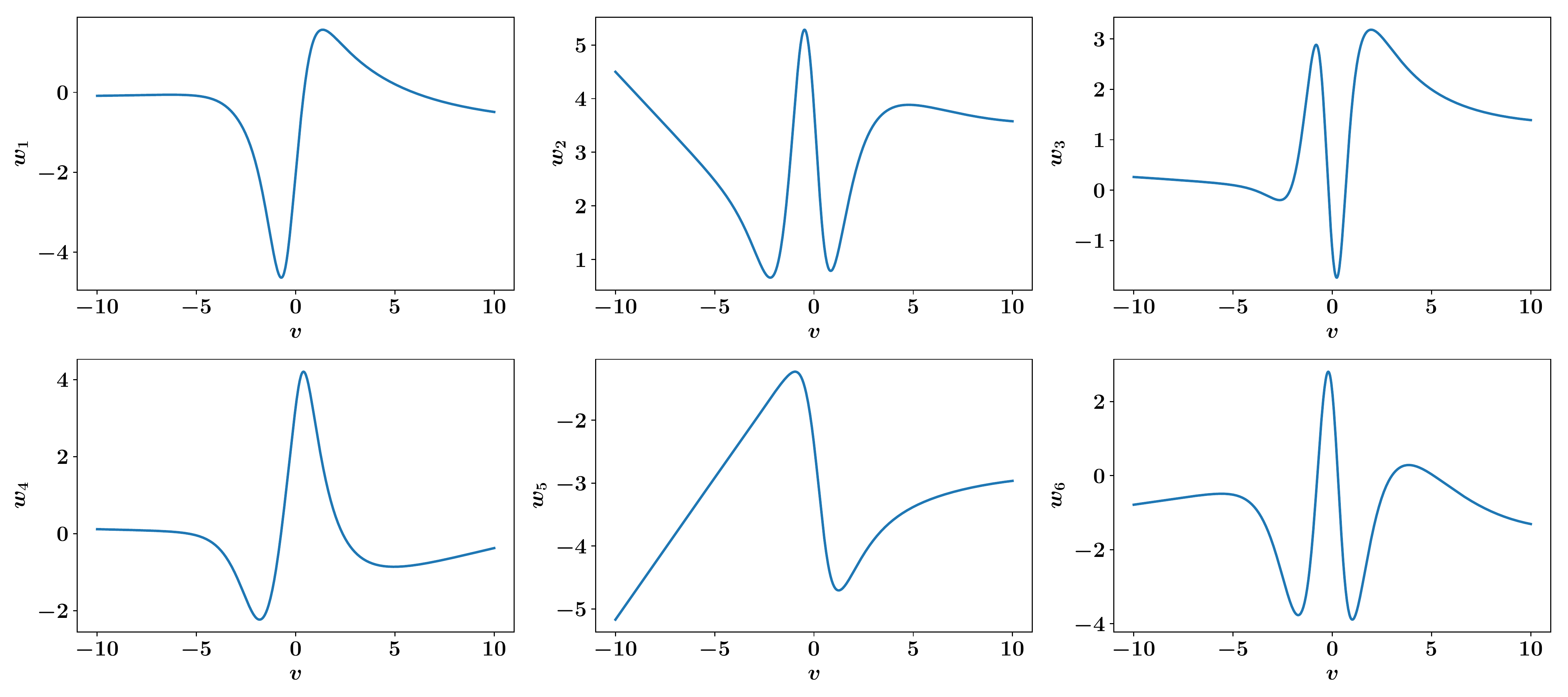}
    \caption{Optimized encoder $\bw(\bv)$ in GalEncMLC, as the generalized moment functions in \taskb for the 1-D BGK model.
}
    \label{fig:mmshapes}
\end{figure}

Fig.~\ref{fig:mmshapes} plots all six generalized moment functions obtained in GalEncMLC for  \taskb for the 1-D BGK model. As we can see they are all well-behaved functions.

\begin{figure}[ht]
\centering
\includegraphics[width=0.98\textwidth]{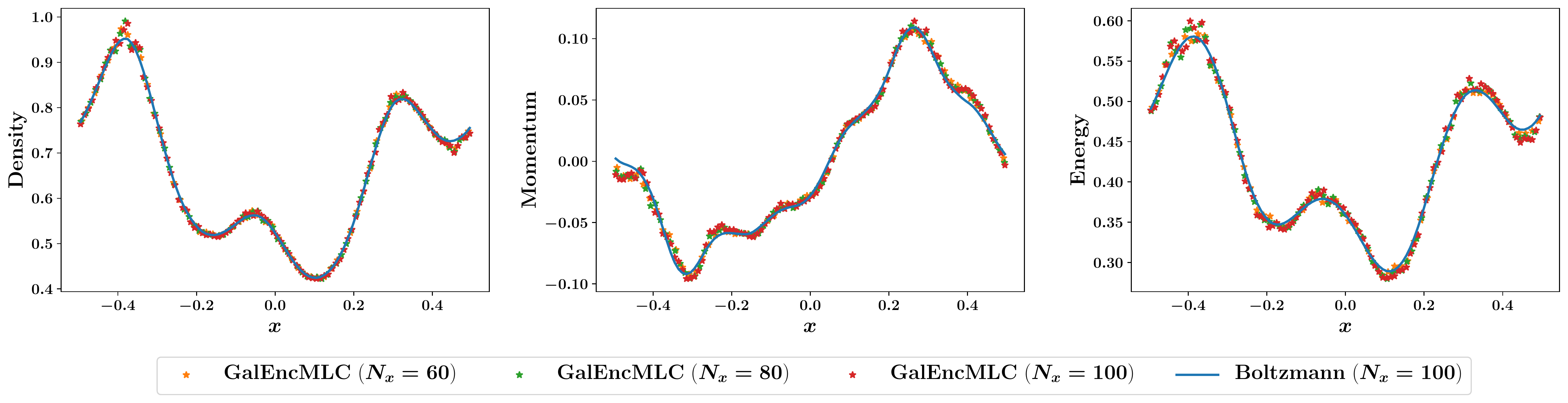}
\caption{The solutions of GalEncMLC at $t=0.1$ with the same initial profile as in \taska but different spatial discretization. The collision kernel is the 2-D Maxwell model and $\veps=8.57$.}
\label{fig:nx_wave_boltz}
\end{figure}

\begin{figure}[ht]
\centering
\includegraphics[width=0.98\textwidth]{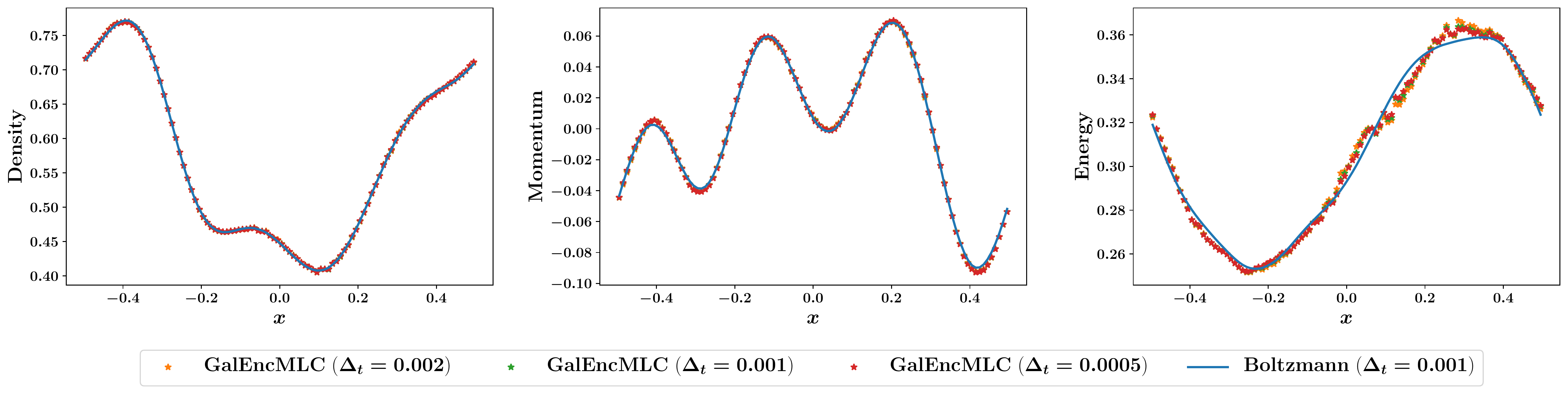}
\includegraphics[width=0.98\textwidth]{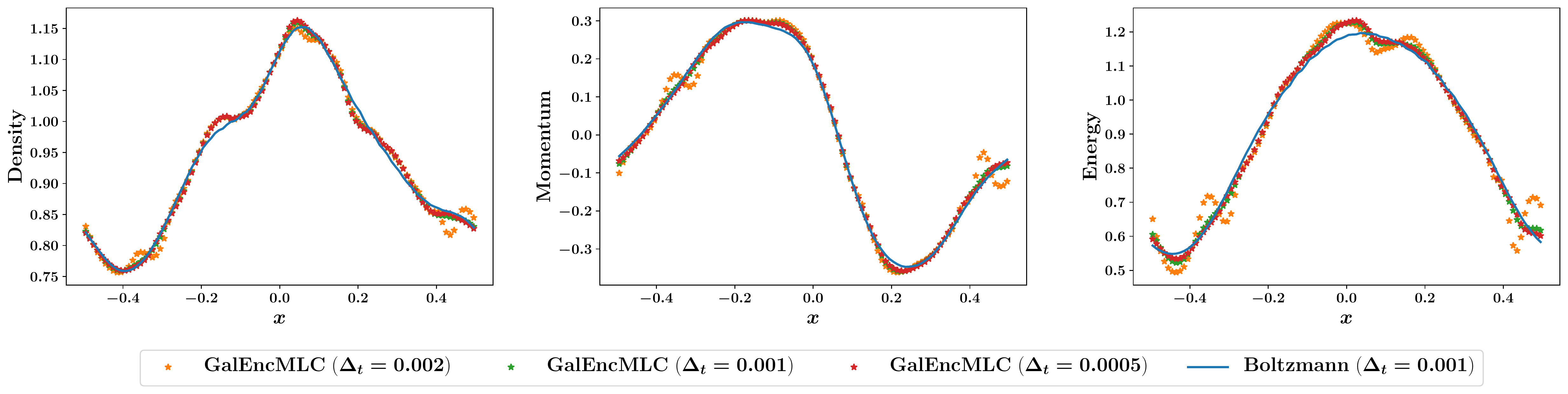}
\caption{The solutions of GalEncMLC at $t=0.1$ with the same initial profile but different temporal discretization. Top: \taska for the 2-D Maxwell model and $\veps=9.95$; Bottom: \taskb for the 2-D Maxwell model and $\veps=6.51$.}
\label{fig:nt_boltz}
\end{figure}

As already discussed in the main text, the machine learning-based moment system behaves like conventional PDEs and are adaptive to different spatial and temporal discretization.  Fig.~\ref{fig:nx_wave_boltz} and Fig.~\ref{fig:nt_boltz} provide more evidence of this point.

\begin{figure}[ht]
\centering
\subfloat{\includegraphics[width=0.46\textwidth]{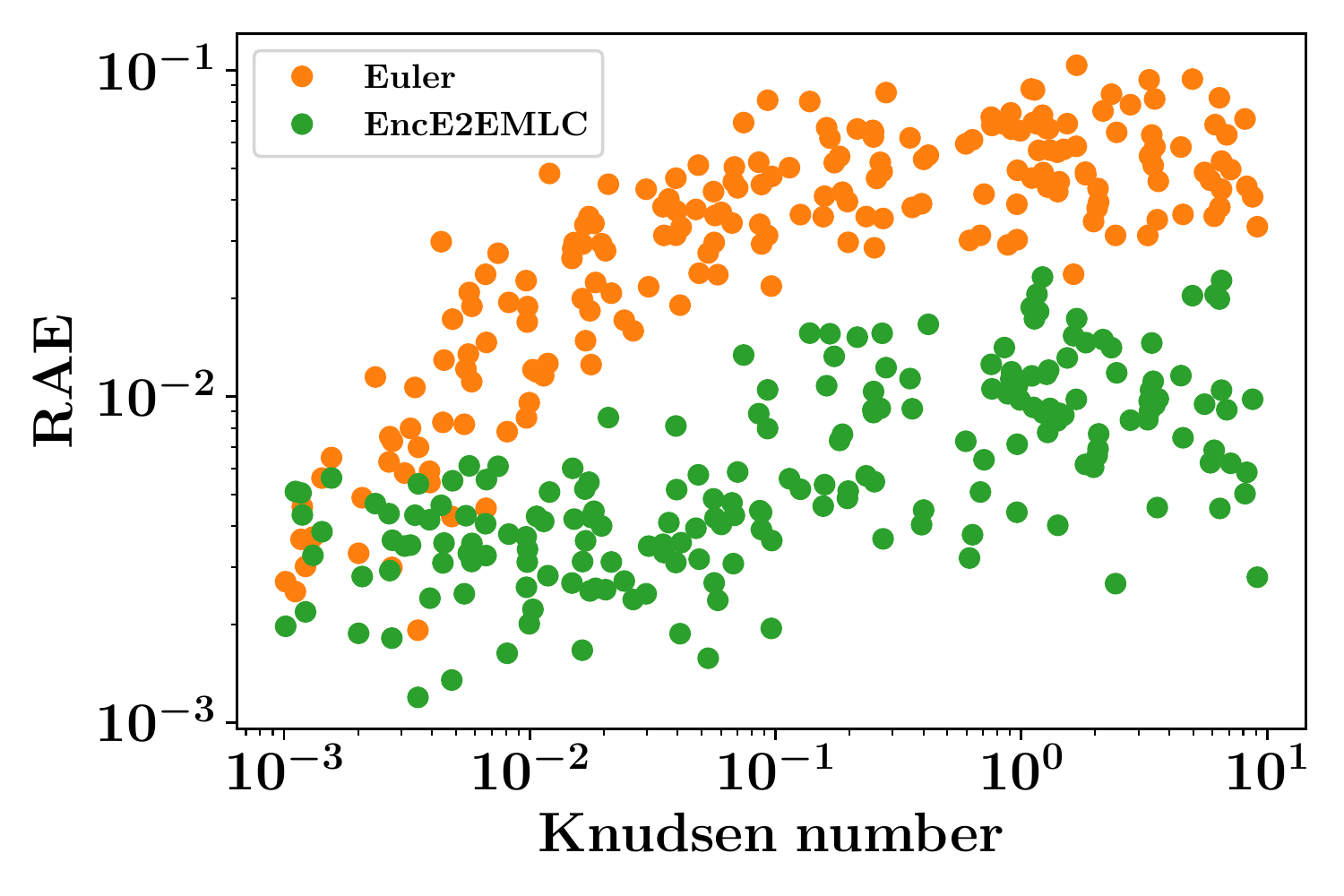}}
\subfloat{\includegraphics[width=0.46\textwidth]{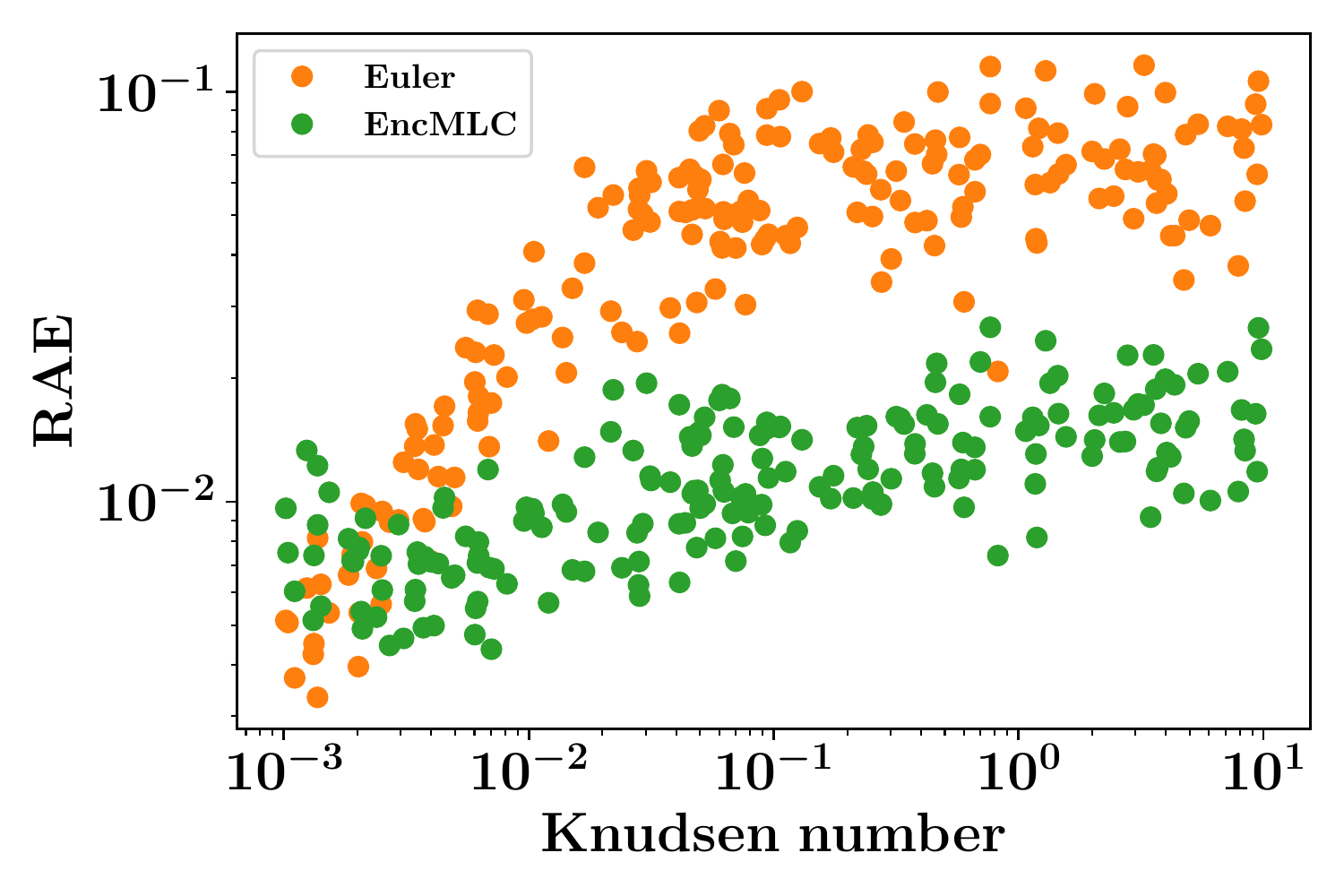}}
\caption{The log-log scatter plot of the relative  error versus the Knudsen number $\veps$ for 200 testing paths in \taska (left) and \taskb (right) for the 1-D BGK model.}
\label{fig:scatter_bgk}
\end{figure}

Fig.~\ref{fig:scatter_bgk} shows the log-log scatter plots of the relative  error versus the Knudsen number $\veps$ for both \taska and \taskb for the 1-D BGK model. 

\begin{figure}[ht]
\captionsetup[subfigure]{labelformat=empty}
\centering
\subfloat{\includegraphics[width=0.33\textwidth]{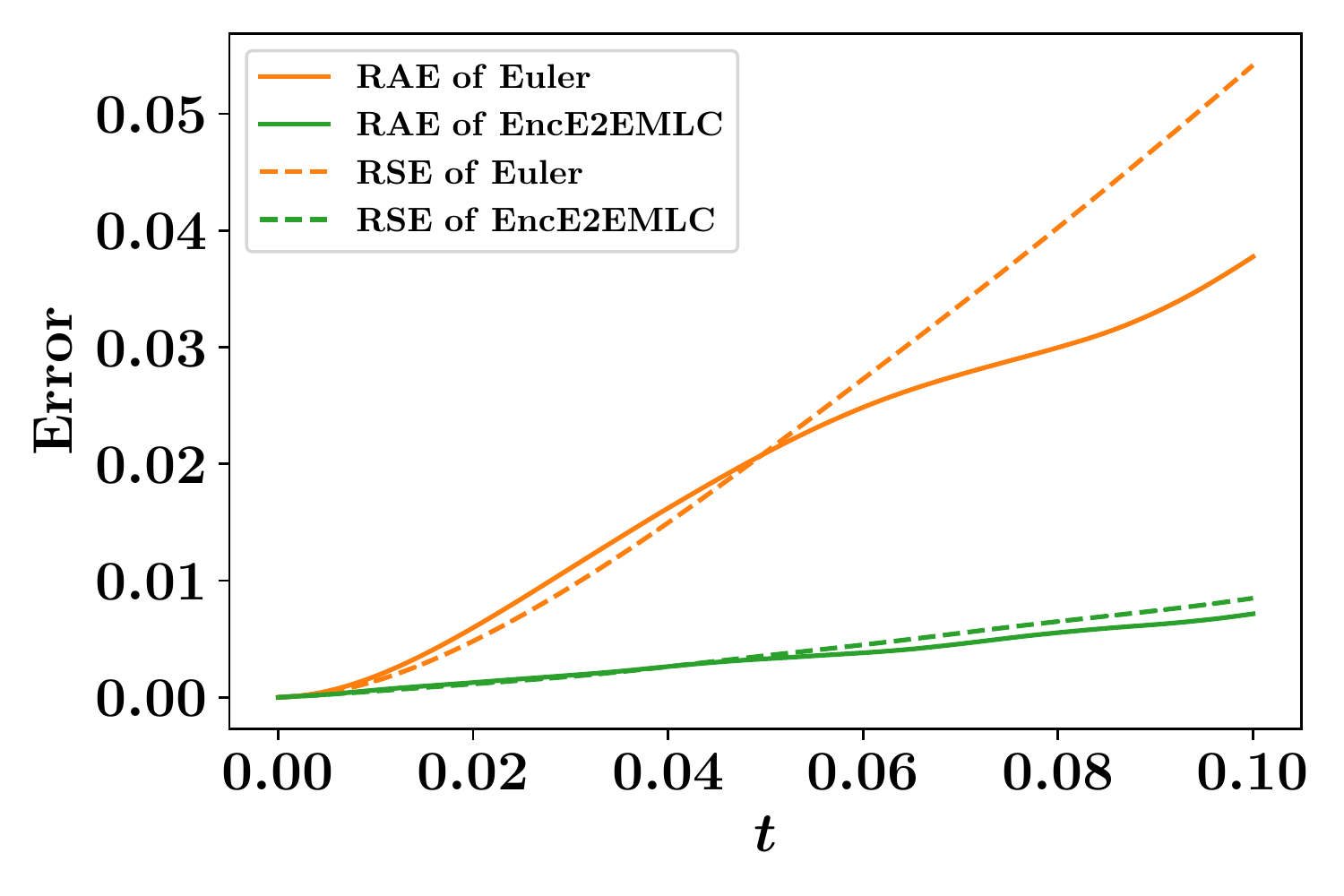}}
\subfloat{\includegraphics[width=0.33\textwidth]{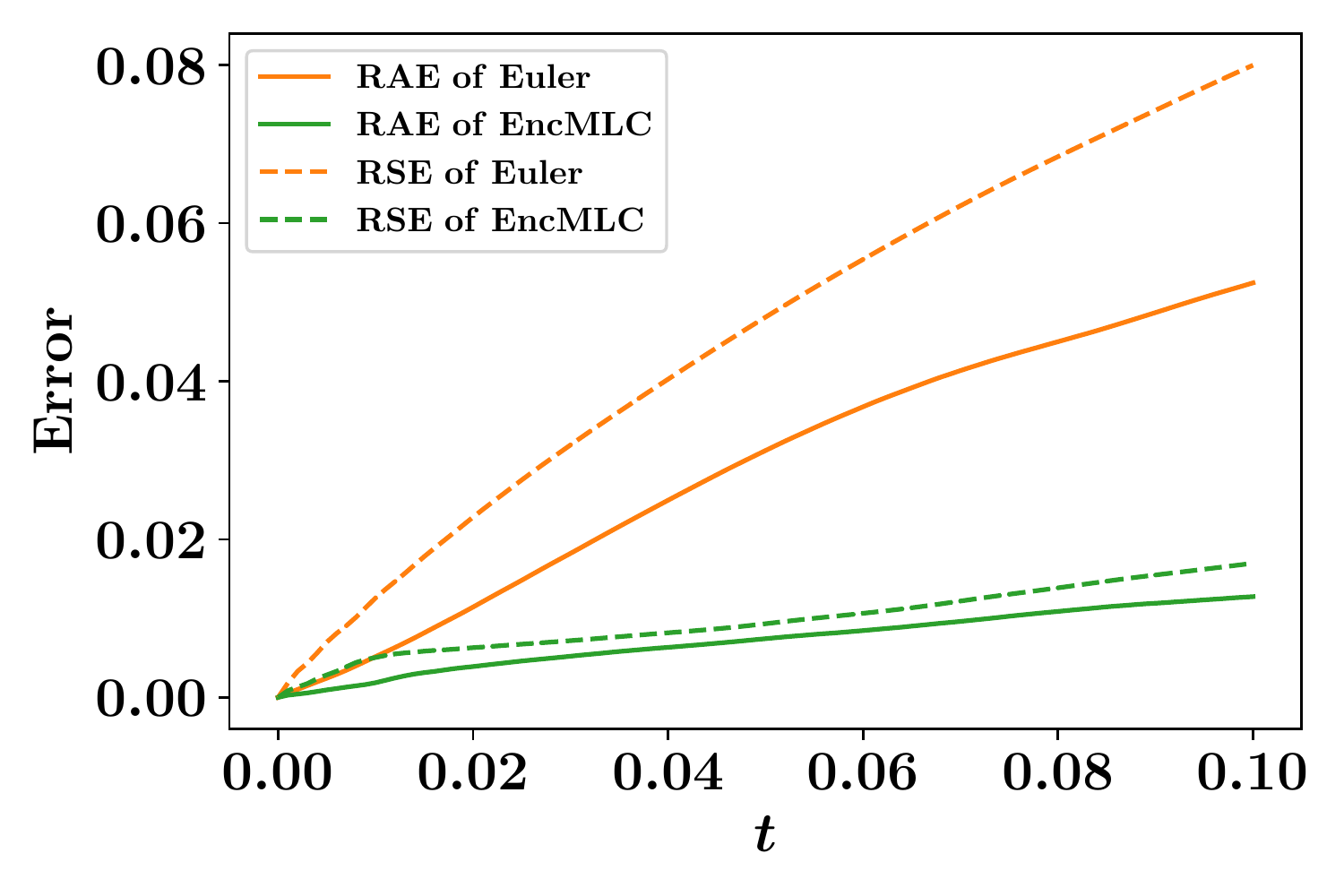}}
\subfloat{\includegraphics[width=0.33\textwidth]{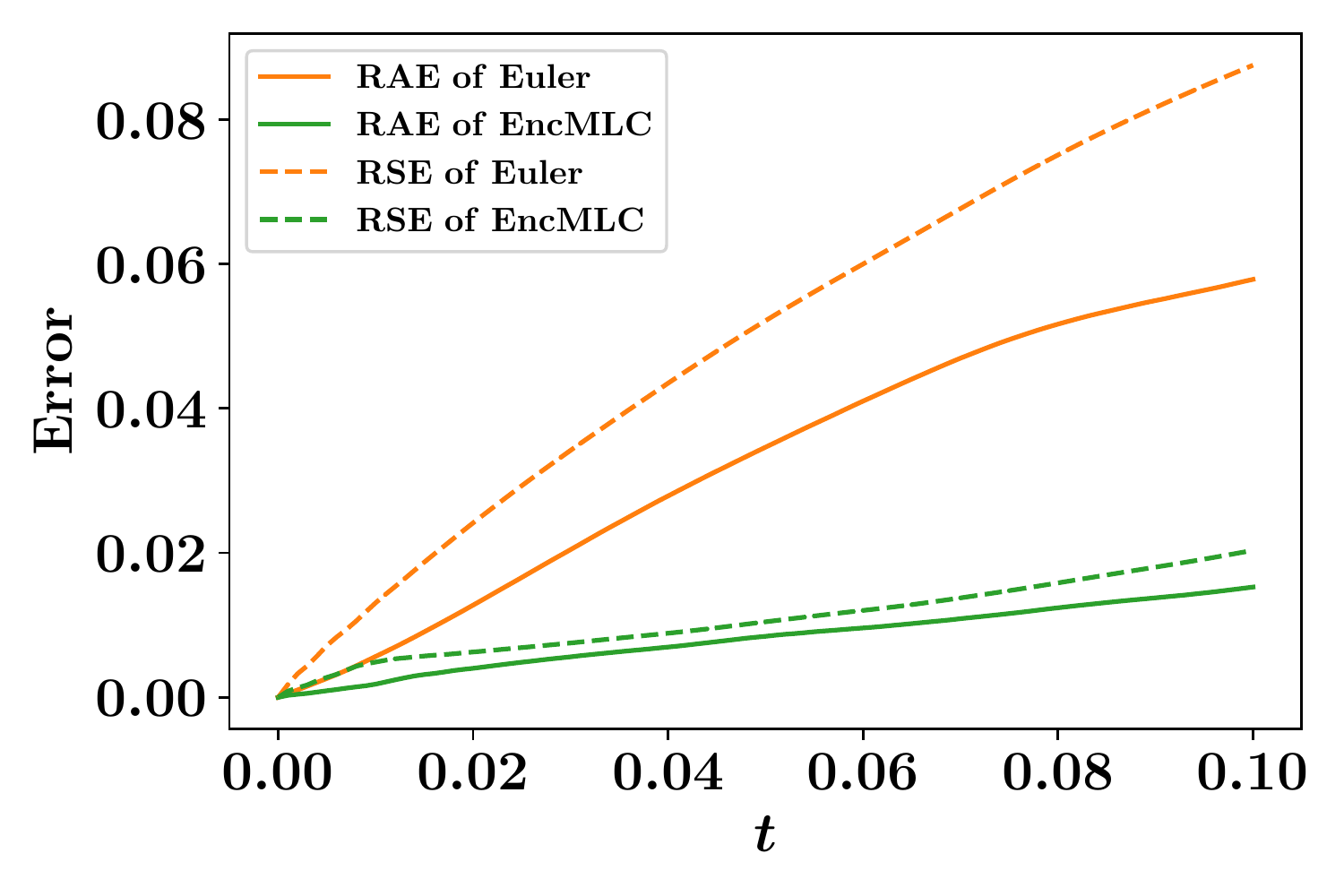}}\\
\subfloat[\textit{Wave}]{\includegraphics[width=0.33\textwidth]{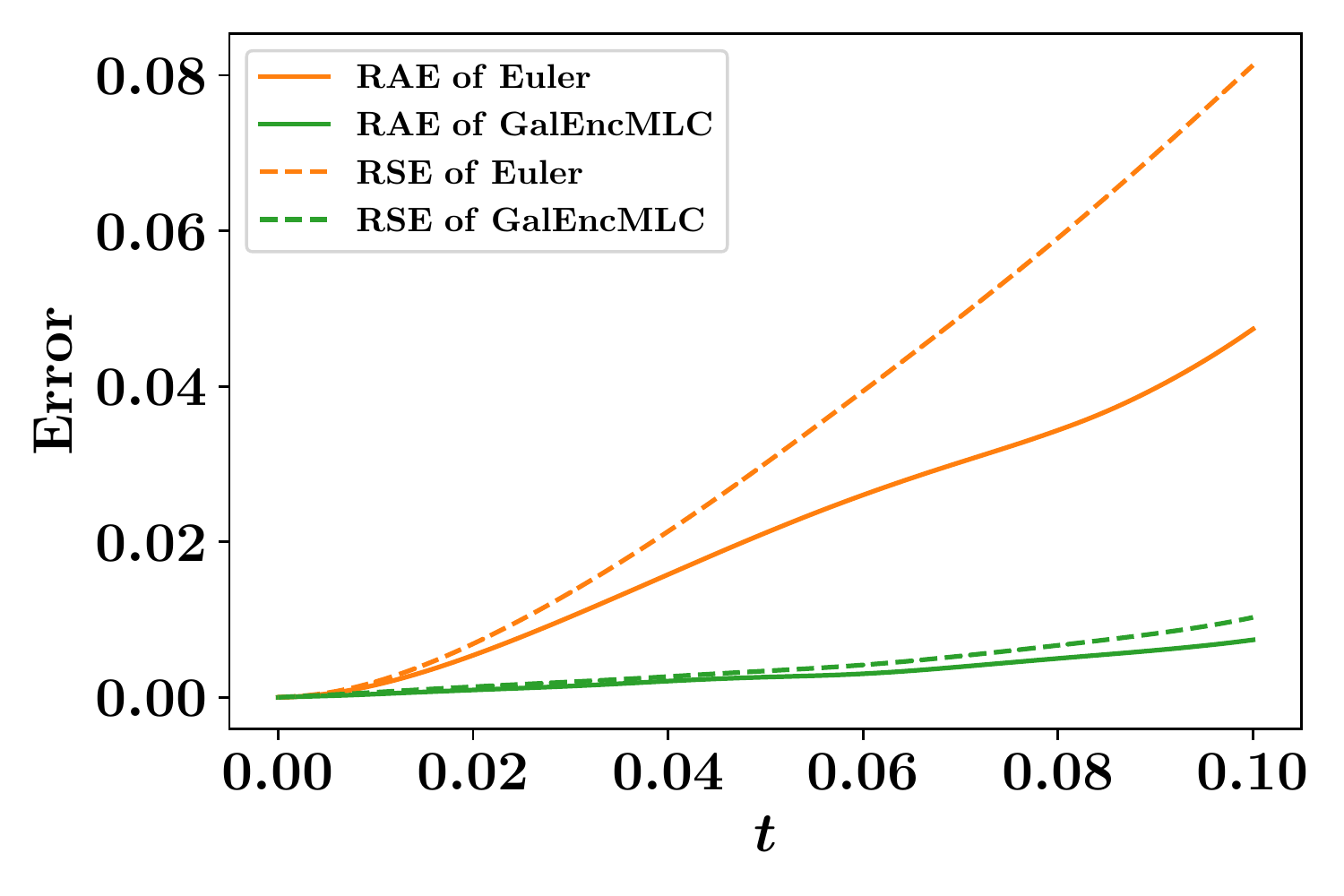}}
\subfloat[\textit{Mix}]{\includegraphics[width=0.33\textwidth]{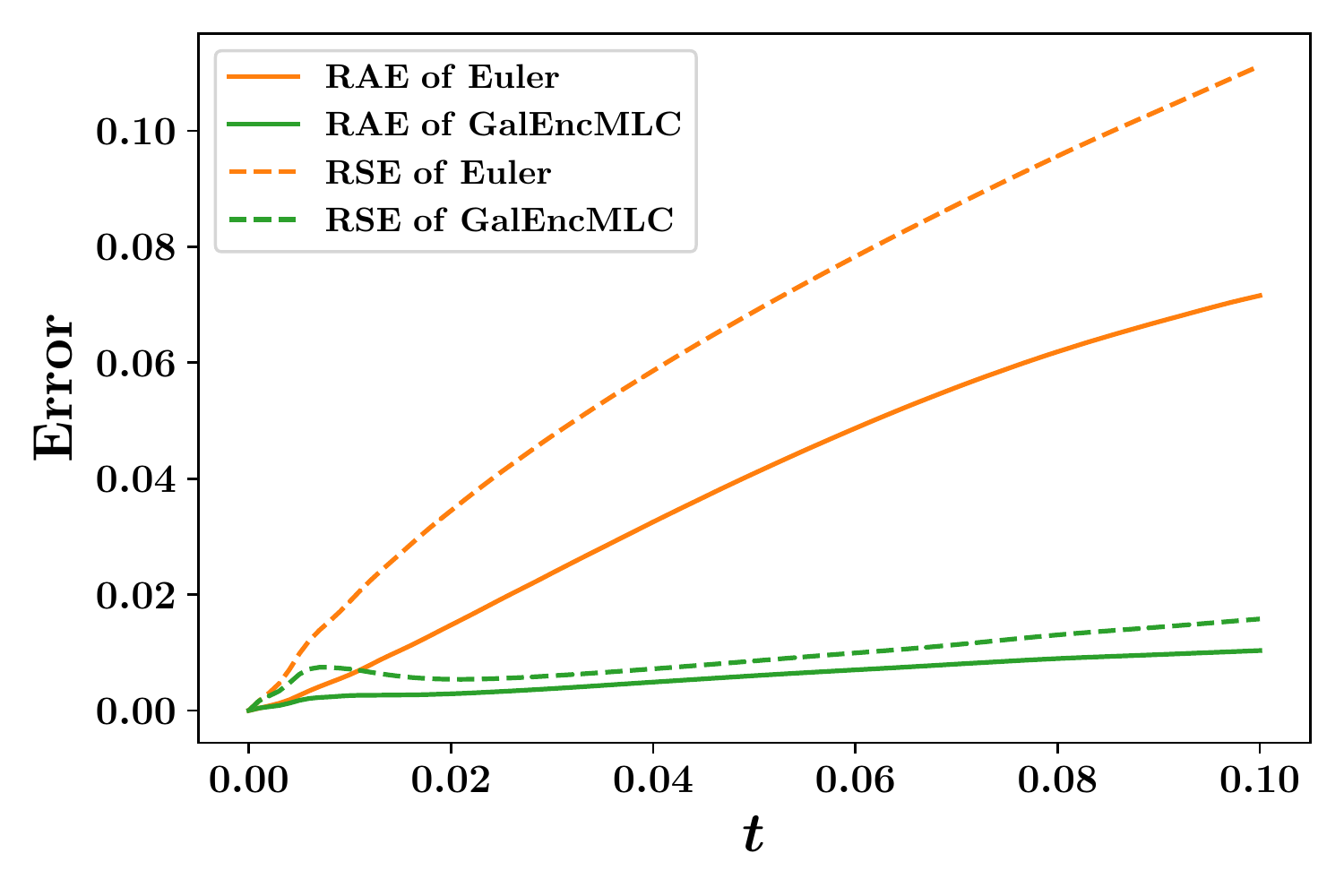}}
\subfloat[\textit{Mix In Transition}]{\includegraphics[width=0.33\textwidth]{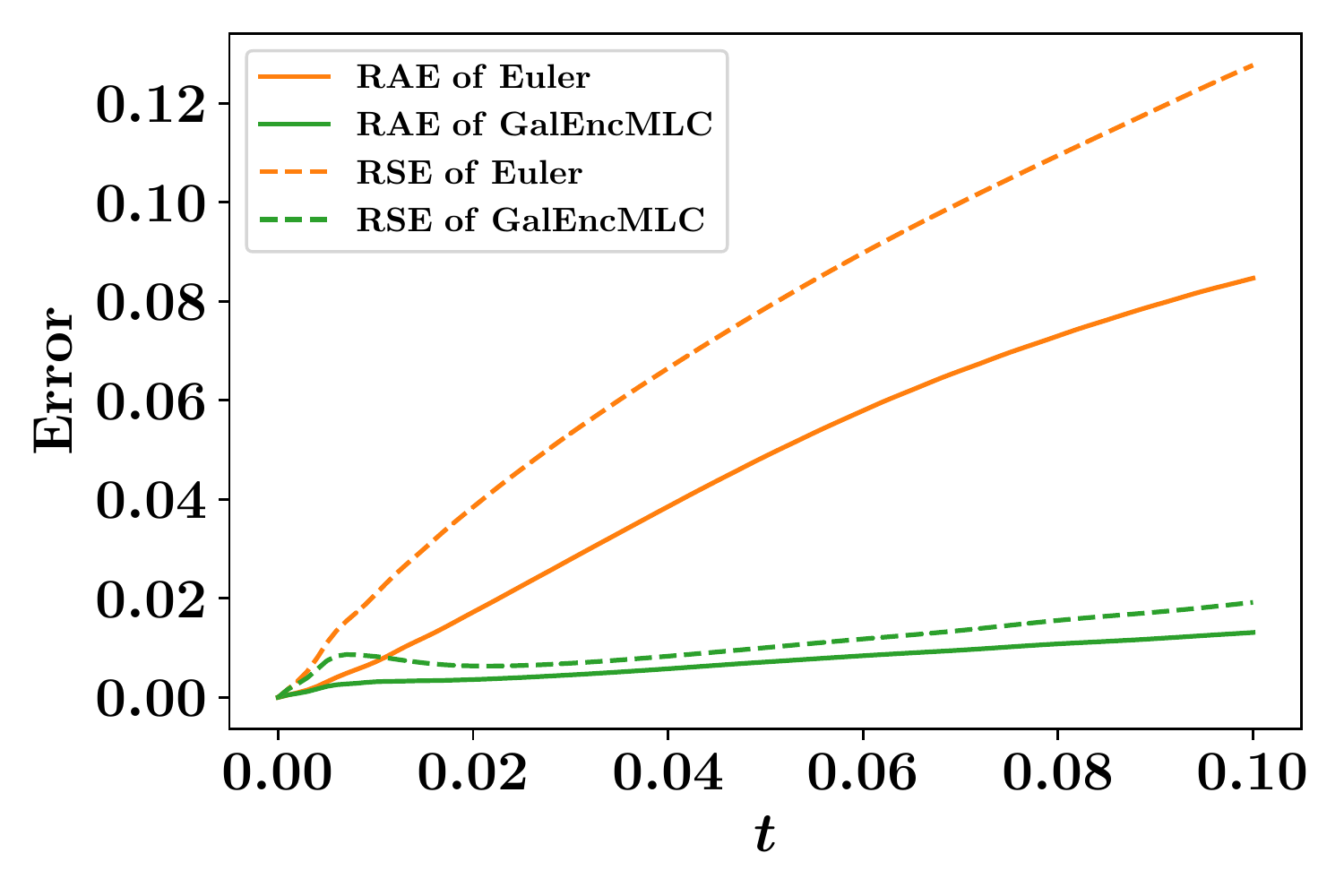}}
\caption{The growth of the relative error as a function of time for 200 new paths in \taska, \taskb, and \taskc. Top: 1-D BGK model; Bottom: 2-D Maxwell model.}
\label{fig:err_curves}
\end{figure}

Fig.~\ref{fig:err_curves} shows the growth of the relative error for the solutions of the Euler equations and machine learning-based moment system on 200 new paths in \taska, \taskb, and \taskc for the 1-D BGK  and 2-D Maxwell model.

\begin{figure}[ht]
    \centering
    \includegraphics[width=0.98\textwidth]{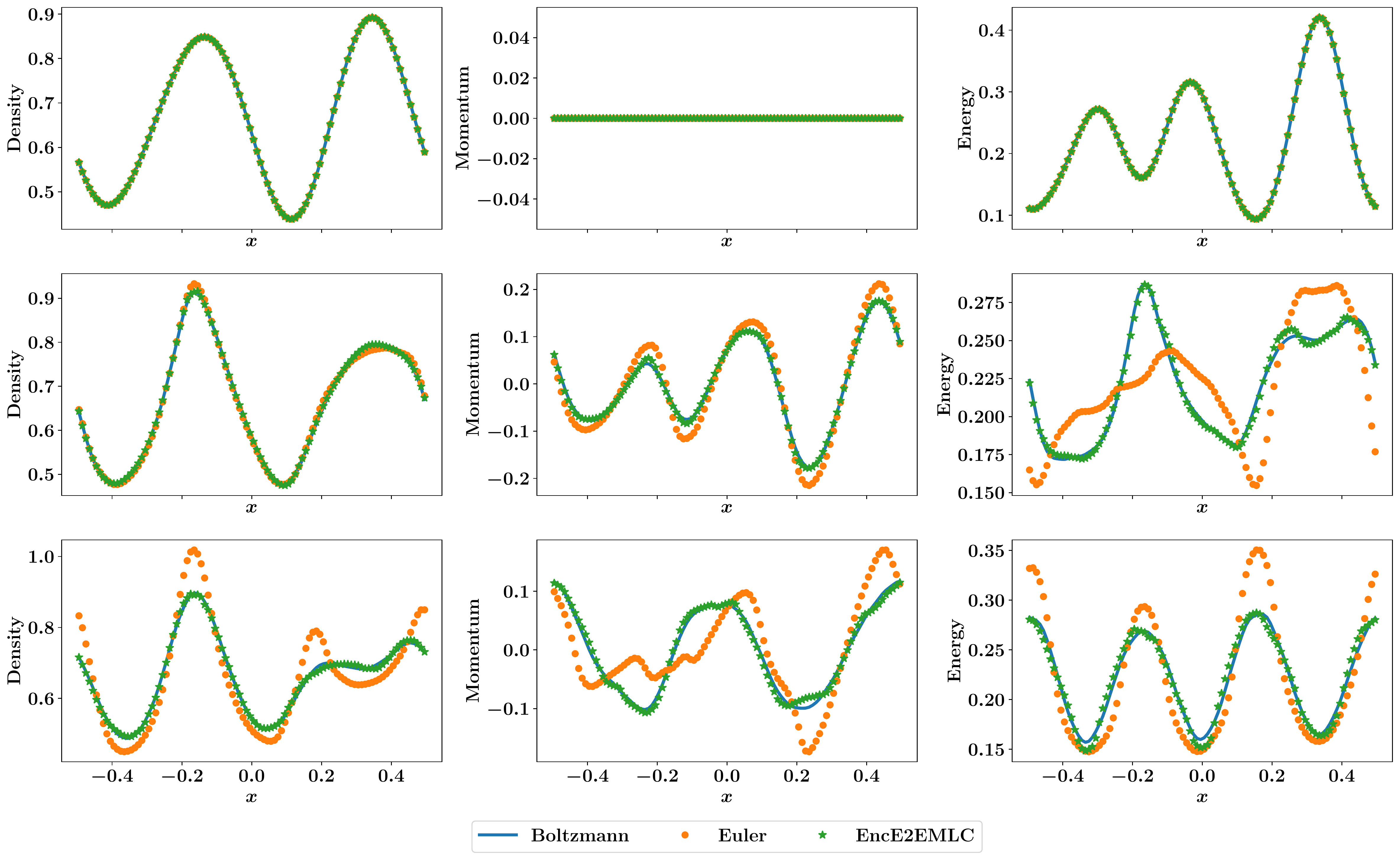}
    \caption{Sample profiles of mass, momentum, and energy densities (from left to right) at $t=0, 0.05, 0.1$ (from top to bottom) in \taska for the 1-D BGK model and $\veps=9.32$, obtained from the kinetic  equation, the Euler equations, and EncE2EMLC.}
    \label{fig:prof_wave_bgk}
\end{figure}

\begin{figure}[ht]
    \centering
    \includegraphics[width=0.98\textwidth]{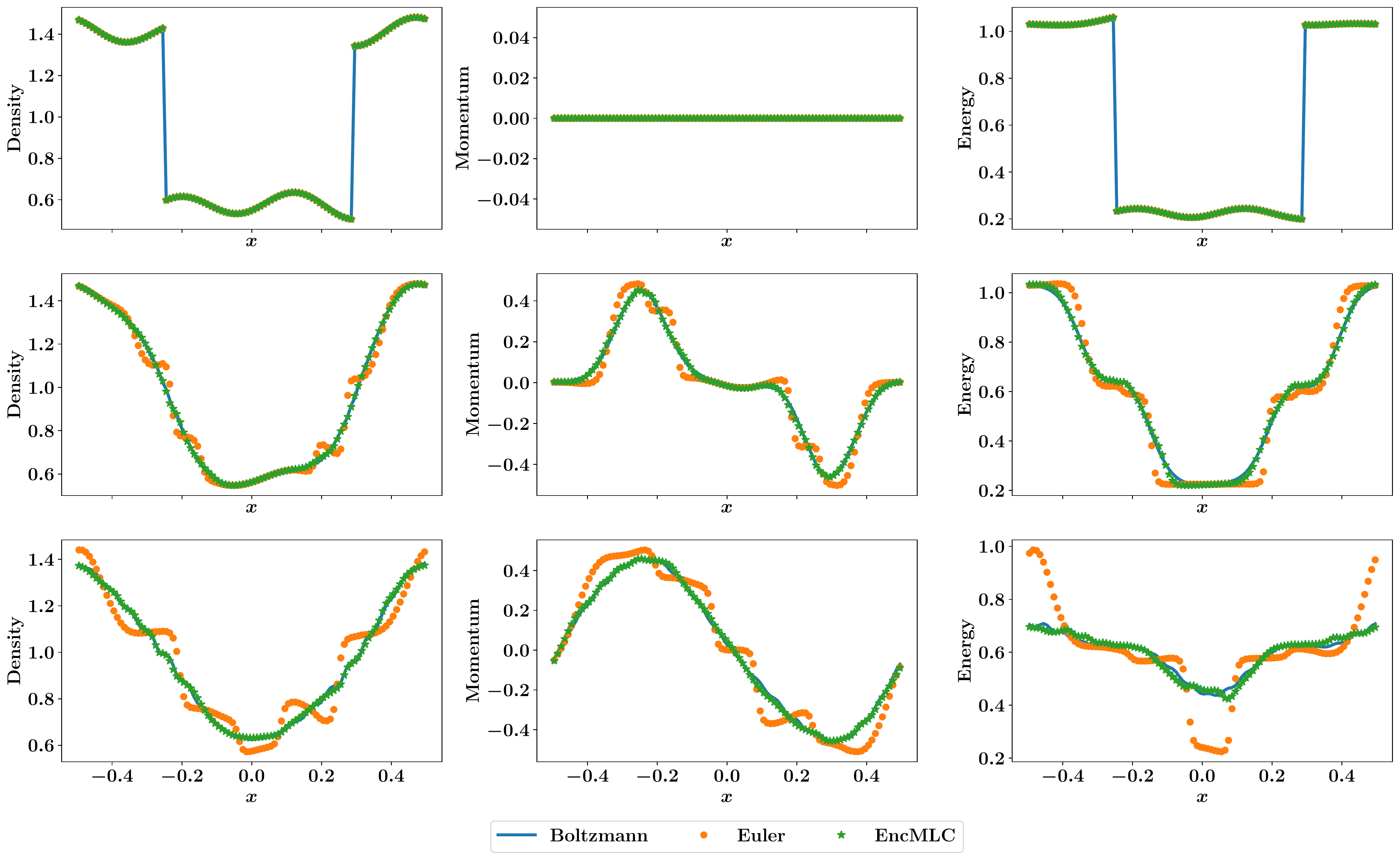}
    \caption{Sample profiles of mass, momentum, and energy densities (from left to right) at $t=0, 0.05, 0.1$ (from top to bottom) in \taskb for the 1-D BGK model and $\veps=5.45$, obtained from the kinetic equation, the Euler equations, and EncMLC.}
    \label{fig:prof_mix_bgk}
\end{figure}

\begin{figure}[ht]
    \centering
    \includegraphics[width=0.98\textwidth]{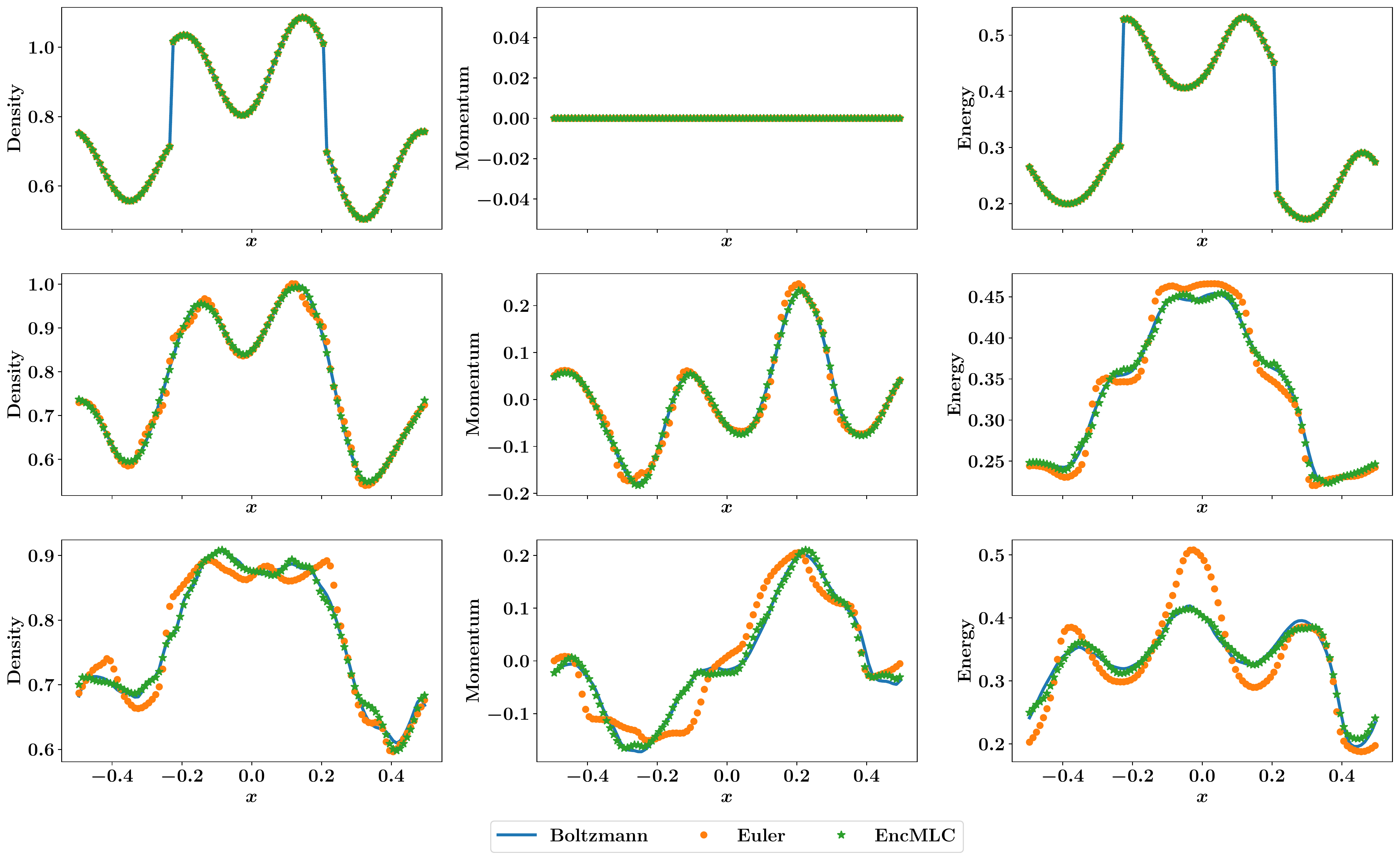}
    \caption{Sample profiles of mass, momentum, and energy densities (from left to right) at $t=0, 0.05, 0.1$ (from top to bottom) in \taskc for the 1-D BGK model, obtained from the kinetic equation, the Euler equations, and EncMLC.}
    \label{fig:prof_transition_bgk}
\end{figure}

Fig.~\ref{fig:prof_wave_bgk}--\ref{fig:prof_transition_bgk} illustrate three sample profiles of mass, momentum, and energy densities in  \taska, \taskb, and \taskc for the 1-D BGK model, obtained  from the kinetic equation, the Euler equations, and the machine learning based moment systems.

\begin{figure}[ht]
    \centering
    \includegraphics[width=0.98\textwidth]{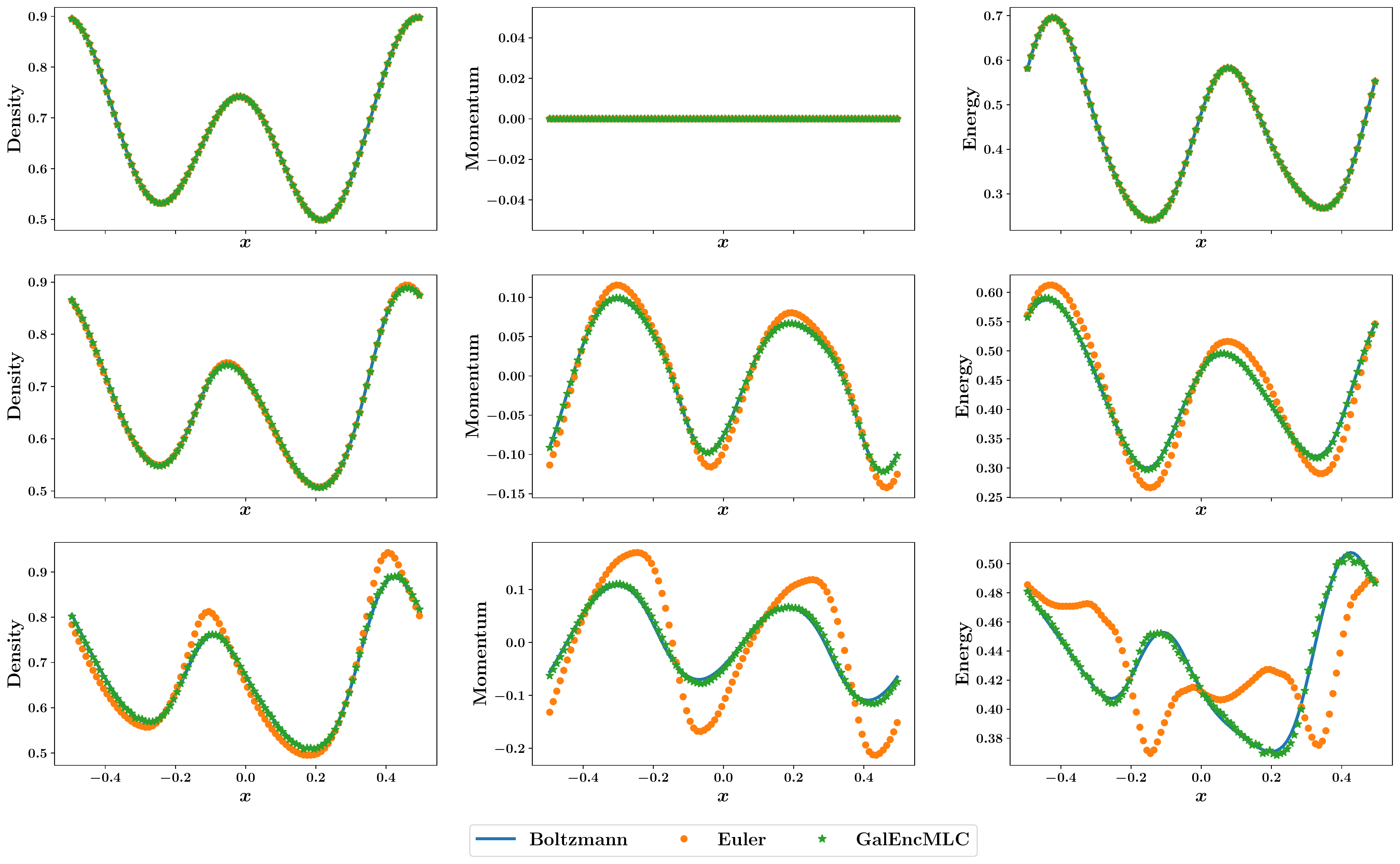}
    \caption{Sample profiles of mass, momentum, and energy densities (from left to right) at $t=0, 0.05, 0.1$ (from top to bottom) in \taska for the 2-D Maxwell model and $\veps=8.09$, obtained from the kinetic equation, the Euler equations, and GalEncMLC.}
    \label{fig:prof_wave_boltz}
\end{figure}

\begin{figure}[ht]
    \centering
    \includegraphics[width=0.98\textwidth]{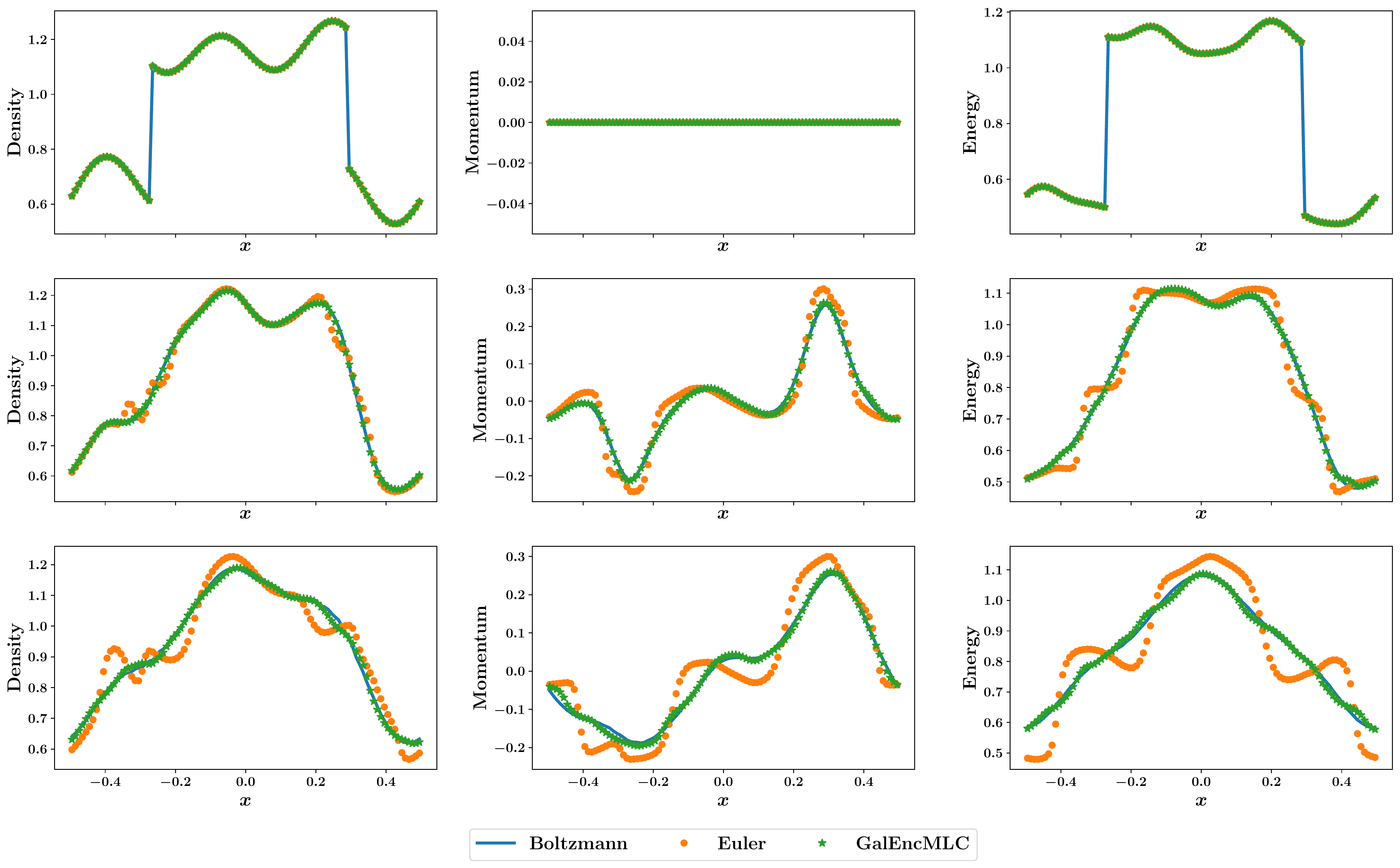}
    \caption{Sample profiles of mass, momentum, and energy densities (from left to right) at $t=0, 0.05, 0.1$ (from top to bottom) in \taskb for the 2-D Maxwell model and $\veps=8.10$, obtained from the kinetic equation, the Euler equations, and GalEncMLC.}
    \label{fig:prof_mix_boltz}
\end{figure}

\begin{figure}[ht]
    \centering
    \includegraphics[width=0.98\textwidth]{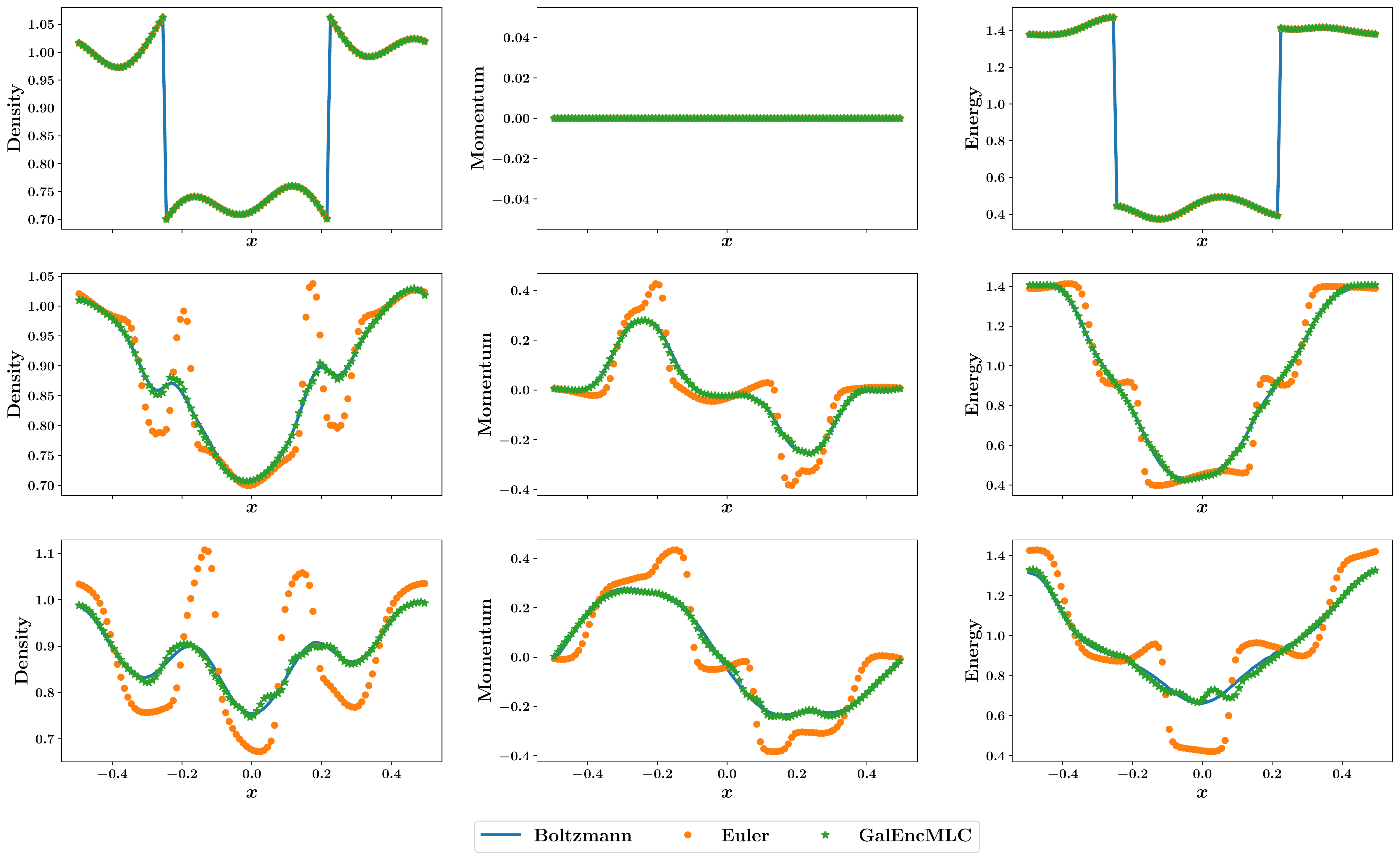}
    \caption{Sample profiles of mass, momentum, and energy densities (from left to right) at $t=0, 0.05, 0.1$ (from top to bottom) in \taskc for the 2-D Maxwell model, obtained from the kinetic equation, the Euler equations, and GalEncMLC.}
    \label{fig:prof_transition_boltz}
\end{figure}

Fig.~\ref{fig:prof_wave_boltz}--\ref{fig:prof_transition_boltz} illustrate three sample profiles of mass, momentum, and energy densities in  \taska, \taskb, and \taskc for the 2-D Maxwell model, obtained  from the kinetic equation, the Euler equations, and the machine learning based moment systems.

\end{document}